\documentclass[useAMS,usenatbib]{mn2e}

\usepackage{graphicx}
\usepackage{times}
\usepackage{natbib}
\usepackage{color}
\usepackage{multirow}
\usepackage{todonotes}
\usepackage{appendix}

\newcommand{\apjs}{ApJS}

\newcommand{\apjl}{ApJL}

\newcommand{\aap}{A \& A}

\newcommand{\enzo}{\it{\small ENZO}}

\begin{document}

\title[Turbulent pressure support and hydrostatic mass-bias in the intracluster medium] {Turbulent pressure support and hydrostatic mass-bias in the intracluster medium}
\author[M. Angelinelli, F. Vazza, C. Giocoli, S. Ettori, T.W.  Jones, G. Brunetti,M. Br\"{u}ggen,  D. Eckert]{M. Angelinelli$^{1,2}$\thanks{E-mail: matteo.angelinelli2@unibo.it}, F. Vazza$^{1,3,4}$, C. Giocoli$^{1,2,5}$, S. Ettori$^{2,5}$, T.W. Jones$^{6}$, G. Brunetti$^{4}$, 
\newauthor
M.  Br\"{u}ggen$^{3}$, D. Eckert$^{7}$ \\
$^{1}$ Dipartimento di Fisica e Astronomia, Universit\'{a} di Bologna, Via Gobetti 92/3, 40121, Bologna, Italy\\ 
$^{2}$ INAF, Osservatorio di Astrofisica e Scienza dello Spazio, via Pietro Gobetti 93/3, 40129 Bologna, Italy\\
$^{3}$ Hamburger Sternwarte, University of Hamburg, Gojenbergsweg 112, 21029 Hamburg, Germany\\
$^{4}$ Istituto di Radio Astronomia, INAF, Via Gobetti 101, 40121 Bologna, Italy\\
$^{5}$ INFN, Sezione di Bologna, viale Berti Pichat 6/2, 40127 Bologna, Italy\\
$^{6}$ University of Minnesota Twin Cities Minneapolis, MN, USA\\
$^{7}$ Astronomy Department, University of Geneva, Ch. d'Ecogia 16, CH-1290 Versoix, Switzerland}

\date{Received / Accepted}
\maketitle
\begin{abstract}
The degree of turbulent pressure support by residual gas motions in galaxy clusters is not well known. 
Mass modelling of combined X-ray and Sunyaev Zel'dovich observations provides an estimate of turbulent pressure support in the outer regions of several galaxy clusters.
Here, we test two different filtering techniques to disentangle  bulk from turbulent motions in non-radiative high-resolution cosmological simulations of galaxy clusters using the cosmological hydro code {\enzo}.  
We find that the radial behavior of the ratio of non-thermal pressure to total gas pressure as a function of cluster-centric distance can be described by a simple polynomial function. The typical non-thermal pressure support in the centre of clusters is $\sim$5\%, increasing to $\sim$15\% in the outskirts, in line with the pressure excess found in recent X-ray observations.
While the complex dynamics of the ICM makes it impossible to reconstruct a simple correlation between turbulent motions and hydrostatic bias, we find that a relation between them can be established using the median properties of a sample of objects. 
Moreover, we estimate the  contribution  of  radial  accelerations to the non-thermal pressure support and conclude that it decreases moving outwards from 40\% (in the core) to 15\% (in the cluster's outskirts). 
Adding this contribution to one provided by turbulence, we show that it might account for the entire observed hydrostatic bias in the innermost regions of the clusters, and for less than 80\% of it at $r > 0.8 r_{200, m}$.
\end{abstract}

\label{firstpage} 

\section{Introduction}
\label{sec:intro}

Turbulence plays a key role in the assembly of large-scale structure and in controlling the physics of the intracluster medium (ICM) \citep[e.g.][]{2014IJMPD..2330007B}. The origin and evolution of turbulence in the ICM have been widely studied using hydrodynamical simulations \citep[e.g.][]{do05,lau09,va11turbo,mi14,2014A&A...569A..67G}.
Various physical processes produce turbulence in galaxy clusters, such as the injection and amplification of vorticity by shock waves \citep[e.g.][]{ry08,2015ApJ...810...93P,va17turbo} or ram pressure stripping \citep[e.g.][]{su06,cassano05, 2007MNRAS.380.1399R}. Moreover, winds from star-burst galaxies and outflows from active galactic nuclei stir the ICM, especially in cluster cores \citep[e.g.,][]{2005ApJ...628..153B,gaspari11a}.

Direct observations of turbulent gas motions in the ICM are almost entirely missing. Only the Soft X-ray Spectrometer (SXS) on board {\sl Hitomi} satellite has directly detected turbulent gas motions in the core of the Persus cluster, a fairly relaxed cluster. Using the width of atomic lines, the root-mean square velocities were found to be $\sim 200 \rm ~km/s$ on $\leq 60 ~\rm kpc$ scales \citep[e.g.][]{hitomi,zuhone18}. This would correspond to a $2-6\%$ non-thermal pressure support in the case of isotropic turbulent motions, or to a $11-13\%$ non-thermal pressure support in case the motions are due to larger-scale sloshing \citep[][]{hitomi18}. \\
Radio observations of Faraday Rotation of polarised sources located behind galaxy clusters hint at a tangled magnetic field in the ICM (with typical coherence scales in the range of $\sim 10-50$ kpc \citep[e.g.][]{mu04,2005A&A...434...67V,bo10}, which is naturally explained by volume-filling stretching motions induced by turbulence \citep[e.g.][]{Dolag:2001,2018SSRv..214..122D,2019MNRAS.486..623D}. 
In order to explain their observed morphology and strength, other indirect probes of turbulent motions are obtained from highly resolved X-ray surface brightness fluctuations, which are interpreted as indications of moderate density fluctuations induced by turbulence \citep[e.g.][]{sc04,2012MNRAS.421.1123C,2014A&A...569A..67G,2014Natur.515...85Z}. From a comparison between X-ray and radio observations, it has been suggested that the surface brightness fluctuations correlate with the diffuse radio emission \citep[][]{eck17,bo18}. This suggests that turbulence detected in X-ray emission could be linked to the re-acceleration of radio emitting particles, via different mechanisms \citep[e.g.][]{bl11b,2020PhRvL.124e1101B}. 
This turbulence is expected also to contribute to the total pressure of the ICM, thus biasing the hydrostatic mass reconstruction \citep[][]{2011MNRAS.416.2567M,Parrish2012,Shi:2014,Shi2015,Shi2016,2018MNRAS.475.1340F,2018PASJ...70...51O,2019arXiv190205420F}. \\
Recently, \citet{eck18published} and \citet{Ettori19} presented results of a systematic study of non-thermal support and hydrostatic mass bias in a sample of galaxy clusters observed for the {\it XMM-Newton} Large Program X-COP \citep[][]{2017AN....338..293E}. The observed mass bias implies that the non-thermal pressure support in the outskirts of nearby, relaxed, massive galaxy clusters (such as the X-COP targets) should vary between 5 to 15$\%$. Such values are a factor of 2 to 3 below what is found in numerical simulations \citep[e.g.][]{lau09,va11turbo,2014ApJ...792...25N,2016ApJ...827..112B,Kay04,Faltenbacher05, Rasia06,ha06,2007ApJ...668....1N}. 
This discrepancy may either be due to missing physics in the simulations such as physical viscosity, magnetic fields, or due to an incorrect separation of turbulent and bulk motions. The results from cosmological simulations may depend on the numerical techniques used to disentangle bulk from turbulent motions. As recently discussed in \citet{2018MNRAS.481L.120V}, different definitions of turbulent motions in numerical simulations could yield non-thermal pressures that differ by factors of 2 to 3, even within the same simulations.
\citet{2019arXiv190207291V} studied turbulent motions in galaxy clusters simulated with (radiative and non-radiative) N-body/SPH codes, using a multi-scale filtering technique. Their results are consistent with \citet{2018MNRAS.481L.120V}, despite the difference in the underlying hydro schemes. This suggests that advanced filtering techniques to study the internal dynamics of the simulated ICM are important to assess the mass-bias in galaxy clusters. \\
In this paper, we measure the non-thermal pressure support by gas motions in the simulated ICM. We apply advanced filtering techniques to identify turbulence in a sample of galaxy clusters produced with high-resolution Eulerian simulations. Our results for the turbulent pressures are then compared to the constraints obtained in the X-COP sample \citep{eck18published}, and in other numerical simulations \citep{2014ApJ...792...25N}.
The paper is structured as follows: in Sect.~\ref{methods}, we describe our cluster sample and the numerical techniques used in the analysis of turbulent motions in the simulated ICM; in Sect.~\ref{sec:res}, we present the results from the analysis of our sample and compare them to recent observational and numerical constraints. In Sect.~\ref{sec:discussion} we analyse some possible sources of discrepancies in the computation of hydrostatic bias, while in Sect.~\ref{sec:conclusions}, we discuss our main findings, the limitations of our analysis and their implications for future work.

\section{Methods}
\label{methods}
\subsection{The Itasca Simulated Cluster sample}
\label{sec:ITASCA}
We used the {\it "Itasca Simulated Clusters"} sample (ISC) for our analysis \footnote{http://cosmosimfrazza.myfreesites.net/isc-project. }, which is  a set of nine galaxy clusters in the $5 \cdot 10^{13} \leq M_{\rm 100}/M_{\odot} \leq 4 \cdot 10^{14}$ mass range simulated at uniformly high spatial resolution with Adaptive Mesh Refinement and the Piecewise Parabolic method in the {\enzo} fluid dynamics code \citep[][]{enzo14}.  Our simulations do not include radiative processes and assumed the WMAP7 $\Lambda$CDM cosmology \citep[][]{2011ApJS..192...18K}, with $\Omega_{\rm B} = 0.0445$, $\Omega_{\rm DM} = 0.2265$, $\Omega_{\Lambda} = 0.728$, Hubble parameter $h = 0.702$,  $\sigma_{8} = 0.8$ and a primordial index of $n=0.961$.  Each cluster was generated from two levels of nested grids as initial conditions (each with $400^3$cells and Dark Matter particles and covering $63^3 ~\rm Mpc^3$ and $31.5^3 \rm~Mpc^3$, respectively). At run time, we also imposed two additional levels of {\it static} mesh refinement in a  $6.3^3 \rm ~Mpc^3$ box around each object, for a fixed $\Delta x = 19.6 ~\rm kpc/cell$ comoving resolution.  More information on the ISC sample can be found in \citet{va17turbo}, \citet{wi17b} and \citet{2018MNRAS.481L.120V}. 

These simulations are non-radiative, in the sense that they do not include radiative gas cooling, nor the effect of heating from star forming regions, reionization or active galactic nuclei. Several studies have shown that the influence of non-gravitational effects is very limited in affecting the global properties of turbulence on the $\gg 100$  $\rm kpc$ scales of interest here, and outside of cluster cores, compared to the impact of mergers and accretion phenomena in the ICM \citep[e.g.][]{va12filter,va13feedback,2019ApJ...874...42V}. However, the combination of cooling and feedback is known to increase the number of density substructures (and in general of the clumping of gas) in the ICM \citep[e.g.][]{nala11,2013MNRAS.432.3030R}, which can locally bias the estimates of gas density and temperature required to compute the global pressure profile of clusters. In this respect, while the gas velocity fields produced by our simulations can be considered realistic enough for regions outside of the cluster core, the gas pressure model of these simulations is likely to be smoother than in reality.

\begin{figure}
\includegraphics[width=0.469\textwidth]{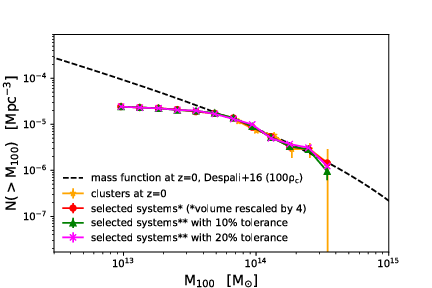}
\caption{Mass functions of selected clusters at different levels of tolerance (i.e. the difference between the expected cosmological mass growth over a given time interval and the measured one). For comparison, the dashed black line gives the theoretical mass function for the cosmology used in our analysis at $z=0$.}
\label{massfunction}
\end{figure}

\subsection{Cosmological selection of independent clusters} \label{sec:cosmoselect}
We take a new approach to building a large sample of galaxy clusters by treating clusters at different redshifts as dynamically independent. Under certain assumptions and for the sake of analysing the properties of turbulent motions in the ICM, these clusters can then be regarded as independent objects \citep{2012MNRAS.422..185G,2016ApJ...818..188D}. Hence, we obtained a sample of 68 clusters from $z\simeq2$ to $z=0$ which are separated in redshift by $\langle \Delta z \rangle  \simeq 0.12 $ that, for the $\Lambda$CDM cosmology, it corresponds to $\langle \Delta t \rangle \simeq 0.91$ Gyr. 
First, we computed r$_{100,c}$ and M$_{100,c}$ of each available snapshot for each object in the $z \leq 1$ range and reconstructed the mass growth of each cluster. Based on this, we could also compute the dynamical time of the cluster in each snapshot, assuming  $t_{\rm dyn} \approx r_{\rm 100}/\sigma_v$, with $\sigma_v = \sqrt{G~M_{\rm 100}/r_{\rm 100}}$, which gives us an estimate for the time between two dynamically independent realizations of the same object. Going back in time from $z=0$, we selected those snapshots that are separated by one dynamical time. \\
Finally, we have to verify that the mass growth between the snapshots is compatible with the expected growth. In particular, we checked that the corresponding $M_{\rm 100}$ mass is {\it below or equal to} the predicted mass, within some tolerance ($0 \div 20\%$), based on the theoretical mass growth for a given $M_{\rm 100}$ at $z=0$ for the given cosmology, as outlined in \citet{2012MNRAS.422..185G} and \citet{2016ApJ...818..188D}. 

\begin{figure}
\includegraphics[width=0.469\textwidth]{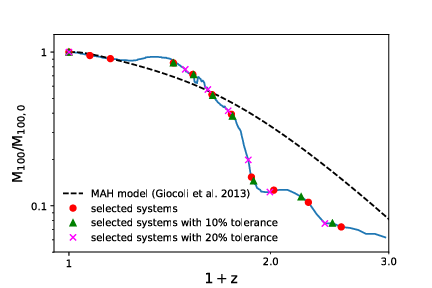}
\caption{Example of real mass growth history (blue solid line) against the theoretical one (black dashed line) for one galaxy cluster of our sample. The different points represent the selected snapshots for different level of tolerance (red points 0\%, green triangle 10\%, pink cross 20\%, see Sect.~\ref{sec:cosmoselect} for details).
\label{figMAH}}
\label{MAHvsmodel}
\end{figure}

We treat each new selected cluster, along the mass growth, 
as independent from the previous one when calculating the 
theoretical mass accretion history \citep{giocoli12b}. In Fig.~\ref{figMAH} the blue curve
displays the mass growth history of one of our cluster from 
$z=0$ to $z=2$. The dashed black line shows the corresponding 
mass accretion history model starting from the $z=0$ system.
The various data points indicate the selected independent 
clusters along the growth with different tolerance thresholds.
Thus, we obtained a final sample of 68 clusters (with $0\%$ tolerance), 
yielding the total mass function shown in Fig.~\ref{massfunction}. 

For comparison, the \citet{despali16} mass function at $z=0$ for the same cosmology and total volume is shown as a black dashed line, 
and this suggests that our final sample is sufficiently mass complete for $M>5\cdot10^{13} M_{\odot}$. 
This  allows us to proceed with a statistical study of the dependence of turbulence on mass, redshift and dynamical state parameters in sub-samples. The limitations connected to this selection procedure are discussed in Sect.~\ref{sec:conclusions}.

\subsection{Identifying turbulence in the ICM}
To disentangle turbulent from bulk motions, we use a small-scale filtering approach. In this technique, we assume that turbulent velocities are approximated as those parts of the gas velocities that fluctuate on the smallest scales, while bulk motions on the largest scales are approximately laminar. The validity of such an approach in cosmological simulations of galaxy clusters is supported by a large body of works on this subject \citep[e.g.][]{do05,lau09,va11turbo,va12filter,mi14,va17turbo}.
With the use of an appropriate small-scale filter, it is possible to define the velocity of the bulk motions and to calculate the velocity of turbulent motions as the difference between the total velocity and the one associated to the bulk motions. In this section, we discuss the updated filtering technique which we used to disentangle turbulent to bulk motions, and the parameters that we tuned to limit the spurious contributions by shocks and clumps.    

\subsubsection{Iterative multi-scale filtering of turbulent motions}
\label{sec:kolmogorov}

The non-thermal to total pressure ratio, $\alpha$, is given by
\begin{equation} \label{eq:alfa}
    \alpha \ \equiv \ \frac{{\rm P_{nt}}}{{\rm P_{tot}}} ,
\end{equation}
where P$_{\rm nt}$ is the non-thermal pressure caused by turbulent motions and P$_{\rm tot}$=P$_{\rm nt}$+P$_{\rm th}$  is the total pressure of the gas.  
P$_{\rm th}$ is the thermal gas pressure, computed as:
\begin{equation} \label{eq:thermal}
\mathrm{P_{th}} = \frac{\mathrm{k_b}}{\mu \ \mathrm{m_p}} \cdot \rho \cdot \mathrm{T} ,
\end{equation}
where $\rho$ is the gas density, T is the gas temperature, k$_{\mathrm{b}}$ is the Boltzmann constant, m$_{\mathrm{p}}$ is the proton mass, $\mu$ is the mean molecular mass for electrons gas and its value is 0.59.

The non-thermal pressure P$_{\mathrm{nt}}$ is estimated as
\begin{equation}
\mathrm{P_{nt}} = \frac{1}{3} \cdot \rho \cdot \delta \rm v^{2}
\end{equation}
where $\delta \rm v$ is the local turbulent velocity; its estimate is in general non-trivial, and in the following we discuss our fiducial procedure to reconstruct it in simulation, as well as test another method used in the literature (see Sec.\ref{sec:nelson}). 

We use an adaptive, iterative filtering to disentangle turbulent from laminar motions in hydrodynamical grid simulations, which follows from previous works by our group \citep[e.g.][]{va12filter,2018MNRAS.481L.120V}. The algorithm does not assume any a-priori coherence scale and the local mean velocity field around each cell is reconstructed with a multi-scale filtering technique, yielding the maximum scale of turbulent eddies by means of iterations in the smoothing scale length. The key assumption is that the gas flow in these simulations is generally part of a cascade of kinetic energy starting from scales much larger than the cell size.

In the original work, we applied a fixed tolerance on the increase of the local rms velocity amplitude with the filtering scale to stop the iterations, and find the smoothing scale of each cell \citep[][]{va12filter}. For a better removal of spurious contribution from shock waves, the method has been later combined with a velocity-based shock finder \citep[][]{va17turbo}.

As a novelty of this work, we apply here instead a more physical definition for the tolerance needed by our iterative algorithm to stop and converge on the local turbulent velocity field. 
In particular, we modify the multi-scale adaptive filtering by \citet{va12filter} to include the scale-dependent expected increase in the local rms velocity. 
In the original work, we applied a fixed tolerance of 1\% to stop the iterations and find the smoothing scale of each cell. Here instead, we adopt a more physical condition and, based on Kolmogorov's theory, we define a variable tolerance $\epsilon_{w}$ for each iteration from the following equation:
\begin{equation}
\label{eq:tolerance}
    \epsilon_{w}=\frac{w^{f}-(w-1)^{f}}{w^{f}} ,
\end{equation}
where $w$ is the size of the smoothing scale in cell's unit and $f$ is the exponent of the Kolmogorov-like relations, which we fix to $0.77$ based on our test, as detailed in the Appendix (Sec.\ref{app:f_test}). At the lower smoothing scale, this value is too high and the best choice is the minimum value between $\epsilon_{w}$ and the fixed tolerance used in \citet{va12filter}. We verified that only for scales smaller than 200 kpc, $\epsilon$ is greater than 1\%.
As discussed in \citet{va12filter}, we define the turbulent velocity in each cell as:
\begin{equation}
\delta {\rm v}  =  {\rm v} -  {\rm v}_{sm}  ,
\end{equation}
where ${\rm v}$ is the velocity field obtained from simulations and ${\rm v}_{sm}$ is the velocity field obtained by a 3D spatial filtering around each i-cell, defined as (in the simple 1D case):
\begin{equation}
{\rm v}_{sm; i} \ = \ \frac{1}{w} \ \Sigma_{j=i-\frac{w}{2}}^{i+\frac{w}{2}} \  {\rm v}_{j}  ,
\end{equation}
where $w$ is the size of the smoothing scale in cell's unit, which determines  the number of cells on which ${\rm v}_{sm}$ is calculated at each iteration step.
We compute the relative variation of the turbulent local velocity $\delta {\rm v}$ between two successive iterations 'w-1' and 'w' as:
\begin{equation}
  \delta_{w} =  \frac{\delta {\rm v}_{w}^2-\delta {\rm v}_{w-1}^2}{\delta {\rm v}_{w}^2} .
\end{equation}

\begin{figure}
\includegraphics[width=0.469\textwidth,height=0.469\textwidth]{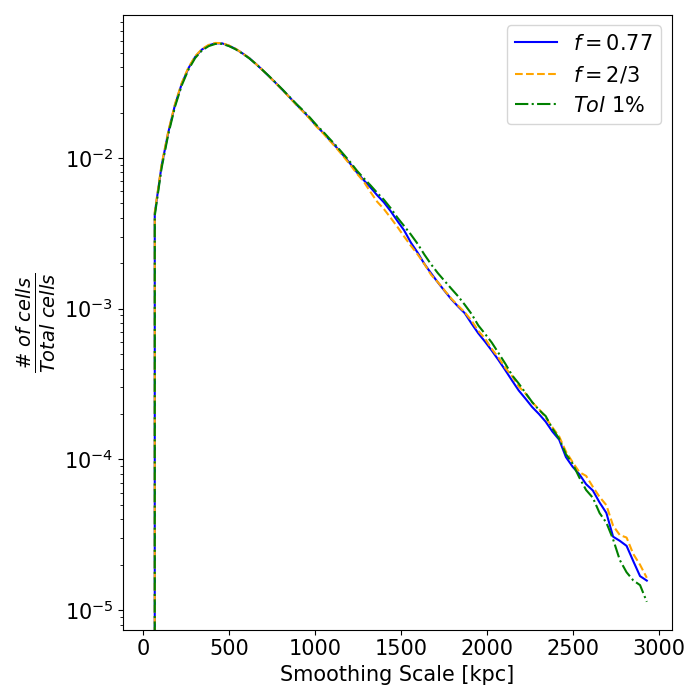}
\caption{Median distribution of smoothing scales for a sub-sample of clusters at $z=0$. The different colors identify different definitions of tolerance, i.e. by varying the exponent for the expected trend of the local rms velocity field as a function of the filtering scale, as explained in Sec.~\ref{sec:kolmogorov}.}
\label{kolmo_scale}
\end{figure}

\begin{figure*}
\includegraphics[width=0.995\textwidth,height=0.4\textwidth]{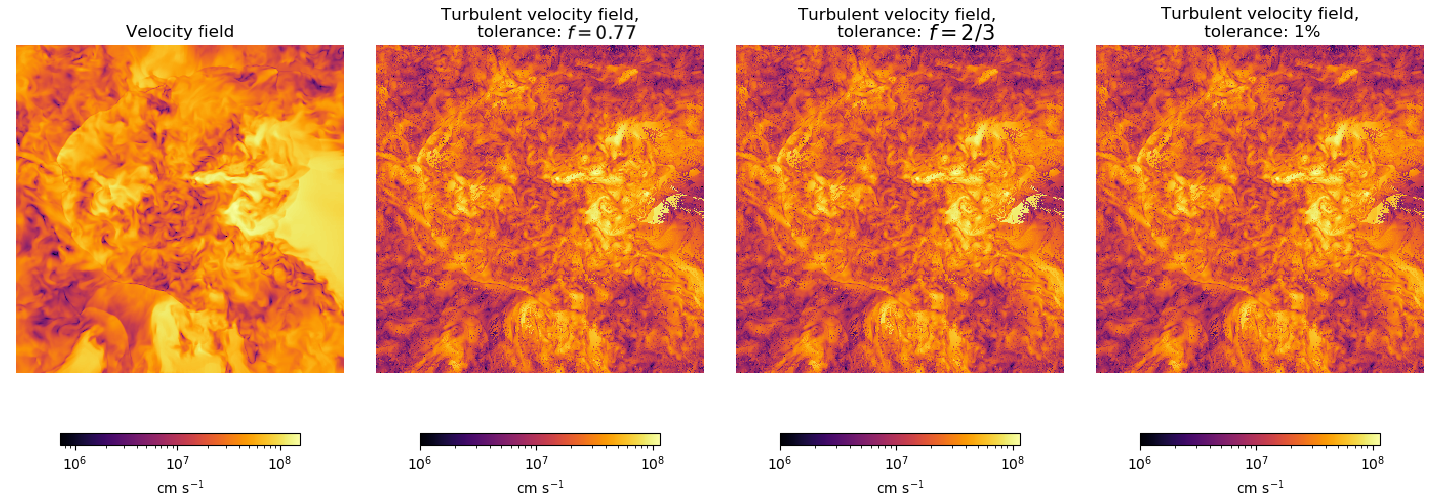}
\caption{Maps of central slice of IT92\_0 at $z=0$. From left to right: First panel: Unfiltered velocity field [cm s$^{-1}$]; Second Panel: Filtered velocity field for tolerance determined by Kolmogorov relation with $0.77$ exponent [cm s$^{-1}$]; Third panel: Filtered velocity field for tolerance determined by the standard Kolmogorov relation [cm s$^{-1}$]; Forth panel: Filtered velocity field for fixed tolerance equal to 1\% [cm s$^{-1}$]}
\label{kolmo_maps}
\end{figure*}

\begin{figure}
\includegraphics[width=0.469\textwidth,height=0.469\textwidth]{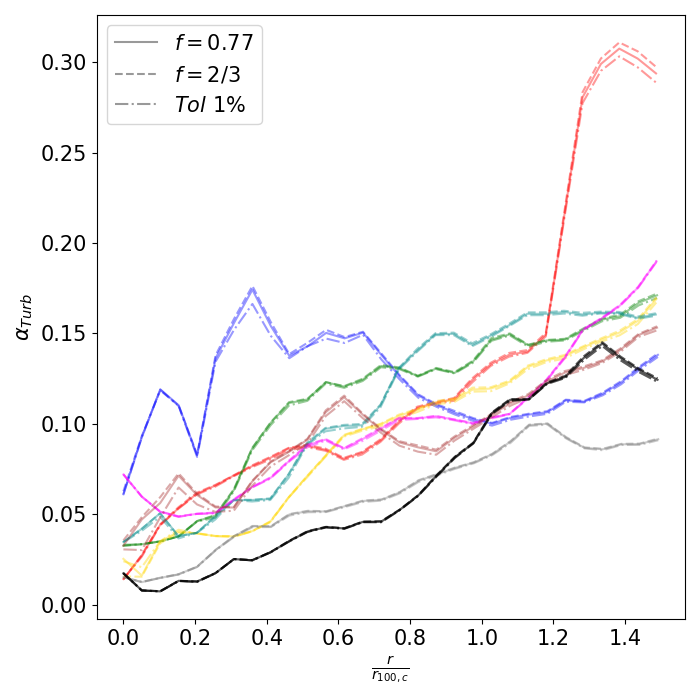}
\caption{Radial profile of non-thermal pressure support for each cluster at $z=0$ for different definitions of the tolerance, $\epsilon_w$ (Eq.\ref{eq:tolerance}) used to stop out iterations on the local turbulent velocity field.}
\label{kolmo_alpha}
\end{figure}

Wherever $\delta_{w}<\epsilon_{w}$, we find the value of turbulent velocity and the value of the smoothing scale.
We test this procedure with two different exponents for the definition of the tolerance and also with a fixed tolerance as described in \citet{va12filter}. The distribution of smoothing scales reconstructed by our algorithm is shown in Fig.~\ref{kolmo_scale}. 
The reconstruction of the turbulent velocity before the application of other filtering techniques is shown in Fig.~\ref{kolmo_maps} for different configurations of the filtering.
Both Fig.~\ref{kolmo_scale} and Fig.~\ref{kolmo_maps} show that the definitions of tolerance have a minor effect on the distribution of the scales or the reconstruction of the turbulent velocity field. This behavior is also visible in the radial profile of $\alpha$, as shown in Fig.~\ref{kolmo_alpha}. 

Here, it is clear that variations in the tolerance lead to small effects on the resulting non-thermal pressure. 
We also tested if this new definition of tolerance could affect the radial behavior of the smoothing scales. We noticed an increase in the smoothing scale of $\leq 20\%$ from the centre of the cluster to the outskirts, which also results in an average increase of the non-thermal pressure at most by $\leq 30\%$ \citep[e.g.][]{va12filter}. However, the radial trend of the turbulent pressure support measured in our data (see following Section) is not an artifact of the filtering procedure: when no filtering is applied, the predicted radial increase of non-thermal support from gas motions in our data \citep[][]{2018MNRAS.481L.120V} as well as in other works \citep[e.g.][]{2014ApJ...792...25N} is much steeper.
In the following, we use the variable tolerance referred to the $f= 0.77$ case, and combine this with the additional filtering of shocks and gas clumps, to better disentangle turbulent motions from other small-scale hydrodynamical features. 
\begin{figure}
\includegraphics[width=0.469\textwidth,height=0.469\textwidth]{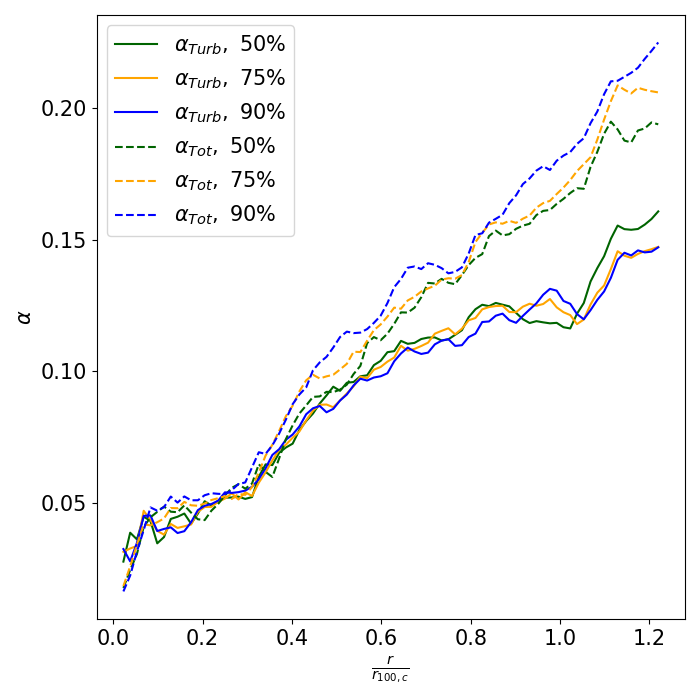}
\caption{Radial profile of median value of non-thermal pressure support for a sub-sample of clusters at $z=0$, obtained by considering the $50\%$, $75\%$ or $90\%$ least dense cells at each radial bin from the centre of clusters. The solid lines are the result for $\alpha_{Turb}$, while the dashed lines are the $\alpha_{Tot}$ ones.}
\label{prof_clump}
\end{figure}

\subsubsection{Radial filtering of turbulent motions} 
\label{sec:nelson}

Other types of filtering techniques are available in the literature.
For the sake of comparison, we also consider the model proposed by \citet[][herefater N14]{2014ApJ...792...25N} to our dataset. 
The definition of the fraction of the non-thermal pressure support in N14 is 
\begin{equation}
  \alpha_{\rm Tot} =  \frac{P_{rand}}{P_{rand}+P_{therm}} = \frac{\sigma_{gas}^{2}}{\sigma_{gas}^{2}+\left( 3 k_{b} T / \mu m_{p} \right)}, 
\end{equation}
where the gas velocity dispersion $\sigma_{gas}$ within radial shells is
estimated as
\begin{equation}
  \sigma_{gas} =  \sqrt{\frac{\sigma_{r}^{2}+\sigma_{t}^{2}}{3}}
\end{equation}
with $\sigma_{r}$ and $\sigma_{t}$ that are the radial and the tangential velocity component, respectively. These components are computed as:
\begin{equation}
  \sigma_{i} =  \sqrt{\left< v_{i}^{2} \right> -\left< v_{i} \right>^{2}} ,
\end{equation}
where $\left< v_{i}^{2} \right>$ is the mean-square gas velocity, while $\left< v_{i} \right>$ is the mean gas velocity, computed in each radial bin, both in radial and tangential direction. Both mean and mean-square gas velocity are weighted by the mass of each gas cell.

In the remainder of the paper we will compare our definition of $\alpha_{\rm Turb}$ to the definition of $\alpha_{\rm Tot}$ above, in which the main difference between them stems from the definition of the "turbulent" velocity. 
In the N14 model, the turbulent velocity is basically the gas velocity dispersion within radial shells, while we filter out also motions which are coherent on small scales (e.g. bulk flows associated to clumps). 
The differences between these types of definition will be discussed with more details in the following sections (for additional discussion see also  \citealt{2018MNRAS.481L.120V}).

\subsubsection{Spurious contributions: shocks and density clumps filtering}

{\it Shock identification} \\
In the study of turbulence, shocks can introduce spurious terms in the estimate of turbulent kinetic energy.
In the presence of shocks, it is possible to use the Rankine-Hugoniot conditions and use velocity or temperature jumps to determine the Mach number. The Mach number is used to calculate the flux of kinetic energy that is dissipated into gas thermal energy.
Here we use the shock finding algorithm based on the the velocity jump between neighbouring cells \citep[][]{va09shocks,va17turbo,2018MNRAS.481L.120V}. 
Detecting shocks with high Mach numbers is relatively easy task in grid simulations with a uniform resolution (and all clusters in the ITASCA sample were simulated with uniform resolution in the "zoom" region), yet the detection of shocks with small Mach numbers is made uncertain by several factors such as numerical errors due to strong gradients or oblique directions of the shocks. In order to reduce the potential noise in the reconstruction of the local turbulent velocity field due to weak shocks sweeping our volume, we set a lower limit to the Mach number of $\mathcal{M}_{\rm thr}=1.3$.  We refer the reader to \citet{va17turbo} and to \citet{2018MNRAS.481L.120V} for an overview of this shock finding method.\\ \\
{\it Clump excision} \\
Dense clumps associated with infalling structures can introduce a bias in the estimate of the local velocity field \citep[e.g.][]{do05}, due to the fact that these structures are correlated with large bulk motions, mostly in the inwards radial direction \citep[e.g.][]{2018MNRAS.481L.120V}. These spurious terms could lead to an overestimate of the non-thermal pressure support. Clumps in simulations are routinely identified as peaks with high density contrast in the radial gas density distribution of the host cluster \citep[e.g.][]{ro11,zhur11}. Therefore, restricting the analysis to a fraction of the gas density distribution at every radius, obtained after excising the highest percentiles in the gas distribution at each radius, is a practical way to limit the bias from the most {\it clumpy} structures in the ICM.
Hence, we tested three different values for masking  the densest cells (considering gas density only) at each radius from the cluster centre: the cells in the top 50\%, 25\% or 10\% of the gas density distribution at every radius. 
As shown in Fig.~\ref{prof_clump}, the profile of non-thermal pressure support $\alpha_{Turb}$ we can derive in our clusters at $z=0$ is overall quite robust against a more restricting selection of cells in the low density part of the distribution at each radius. On the other hand, when we use the same cells selection to compute the radial profile of $\alpha_{Tot}$, we notice a larger impact of the gas clumping factor, shown by the increase of non-thermal support when denser cells are retained in the procedure \citep[see][for a detailed study on the ICM inhomogeneities]{2013MNRAS.428.3274Z}. This suggests that $\alpha_{Tot}$ is susceptible to clump expulsions, while $\alpha_{Turb}$ is more stable respect to the presence of clumps.
Based on our results and previous work \citep[e.g.,][]{2013MNRAS.428.3274Z, 2013MNRAS.432.3030R}, we will use the 90\% masking in our analysis which roughly mimics the approach applied to the X-ray spatial analysis \citep[e.g.,][]{2017arXiv170802954G,eck18published}. 
Our tests show that adoption of ${\mathcal M}\ge 1.3$ and the exclusion of cells in the top 10 percentile in density in each shell yields the best filtering combination. 
In the following, we will refer to the results of our best filtering configuration as the turbulent velocity.

\begin{figure*}
\includegraphics[width=0.995\textwidth]{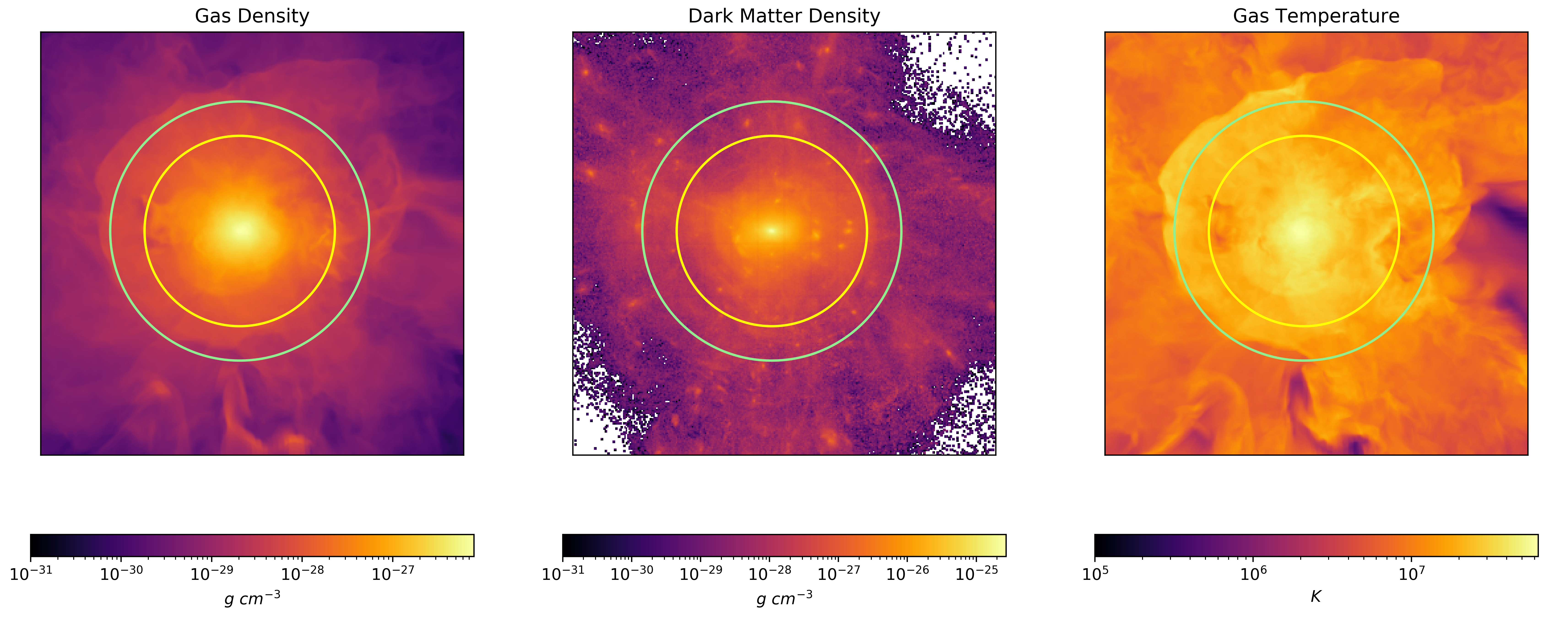}
\includegraphics[width=0.995\textwidth]{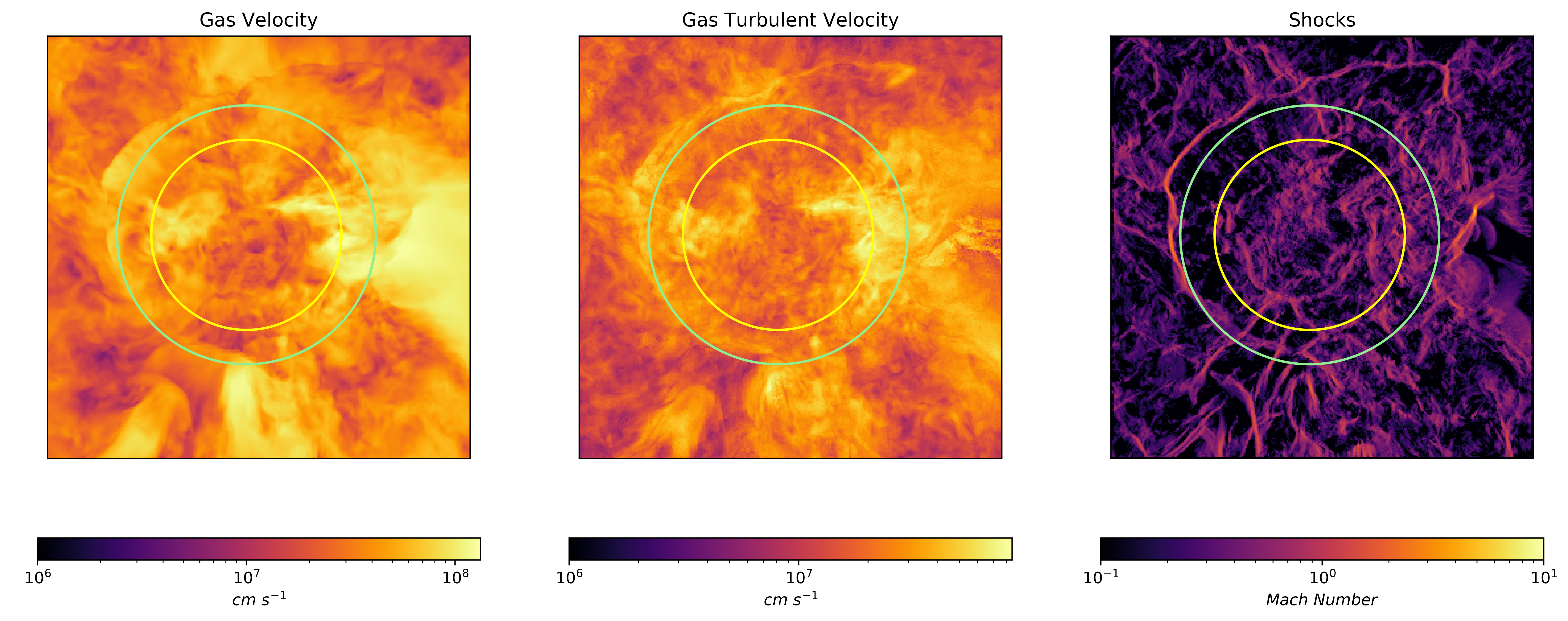}
\caption{Maps of the central region in IT92\_0 at $z=0$. The inner circle is r$_{200,c}$, while the outer one is r$_{100,c}$. From top left to bottom right: First panel: gas density [g cm$^{-3}$]; Second Panel: dark matter density [g cm$^{-3}$]; Third panel: gas temperature [K]; Forth panel: unfiltered velocity field [cm s$^{-1}$]; Fifth panel: turbulent velocity field [cm s$^{-1}$]; Sixth panel: Shocks Mach Number.}
\label{maps_all}
\end{figure*}

\section{Results}
\label{sec:res}

Our projected maps of gas density, dark matter density, gas temperature and unfiltered velocity field, as well as the turbulent velocity field and shocks for one of the objects in our sample are given in  Fig.~\ref{maps_all}.
The maps well illustrate the complex gas flow patterns that are typically found in the simulated clusters at all epochs, with expanding shocks that mark recent heating episodes in the ICM (top and lower right panels) and 
a mixture of large-scale bulk flows (lower left panel) and small-scale turbulent motions (lower central panel). 

To extract the radial profiles of $\alpha$, we first define the cluster center based on the maximum value of the thermal energy of the gas at each snapshot since this definition of the centre results makes the most stable, in highly perturbed systems, as well.

We consider mass-weighted values of the pressures, estimated on the same cells selected with our turbulent filtering technique (see Sec.\ref{sec:kolmogorov}).

\begin{figure*}
$\begin{array}{cc}
\includegraphics[width=0.469\textwidth,height=0.469\textwidth]{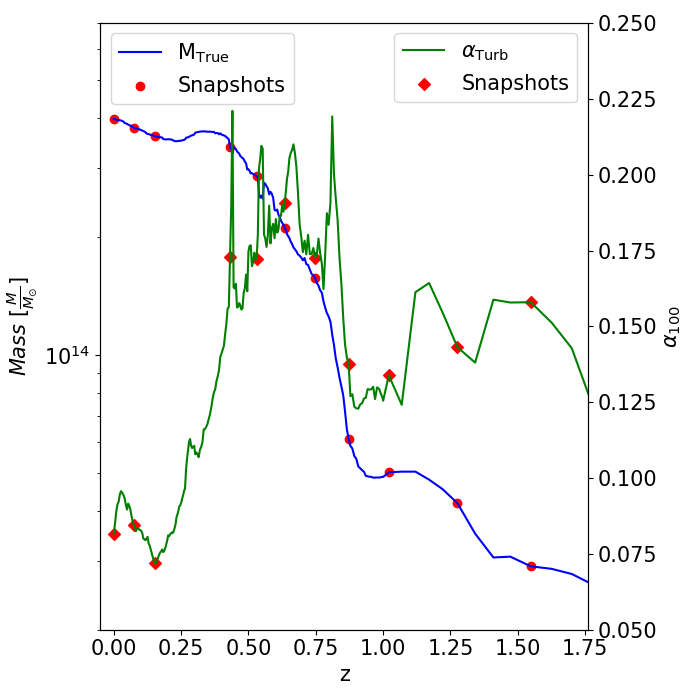}  &
\includegraphics[width=0.469\textwidth,height=0.469\textwidth]{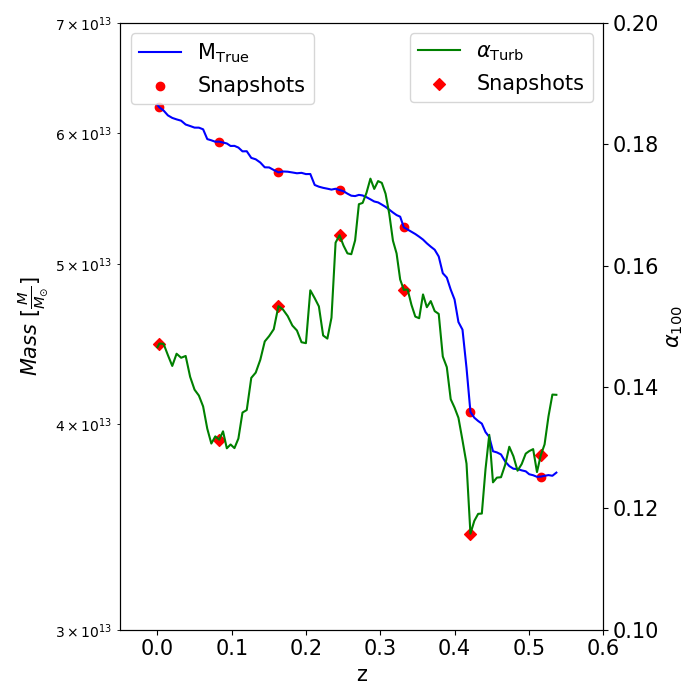}  
\end{array}$
\caption{$M_{100,c}$ growth (blue solid line) and non-thermal pressure support at r$_{100,c}$ time behavior (green solid line) for IT92\_0 and IT90\_4. The red points are the selected snapshots as explain in Sect.~\ref{sec:cosmoselect}.}
\label{mass_alpha_grow}
\end{figure*}

In the following analysis, we consider two methods to estimate the ratio $\alpha=P_{nt}/P_{tot}$:
(i) either applying our filtering technique ($\alpha_{\rm Turb}$) or the N14 model ($\alpha_{\rm Tot}$);
(ii) by comparing the mass estimated from the hydrostatic equilibrium equation with the total mass distribution ($\alpha_{\rm HS, Turb}$ or $\alpha_{\rm HS, Tot}$). 
Furthermore, when we refer to $\alpha$ or $\alpha_{\rm HS}$, we are considering both "Turb" and "Tot" quantities at the same time. Otherwise, if we consider only one of these quantities, we will indicate it with the related pedice.
In Fig.~\ref{mass_alpha_grow}, we show values of mass and non-thermal pressure support $\alpha_{\rm Turb}$ at radius r$_{100,c}$ in function of redshift. We notice that there is a strong relation between mergers and an increase of $\alpha_{\rm Turb}$. Instead, when the cluster is not affected by mergers the value of $\alpha_{\rm Turb}$ decreases. 
The red points in Fig.~\ref{mass_alpha_grow} are the selected snapshots obtained by the selection described in Sect.~\ref{sec:cosmoselect}.
In the following, we will refer to three different typical radii R$_{500,c}$, R$_{200,c}$, R$_{100,c}$ and R$_{200,m}$. From the cluster's center to the peripheries we find R$_{500,c}$:R$_{200,c}$:R$_{100,c}$:R$_{200,m}$ and they are related by these approximated ratios 1:1.4:1.9:3 \citep[see][and references therein]{Walker19}.

\subsection{Parametrising the profile of non-thermal pressure support in galaxy clusters} \label{sec:fitting}

The radial distribution of non-thermal pressure support we find in our cluster sample is so regular that an analytic formula reproduces well the trend  of $\alpha_{\rm Turb}$ with radius: 
\begin{equation} \label{eq:fitalfaturb}
    \alpha_{\rm Turb}(r) \ = \ a_{0} \ \cdot \ \left( \frac{r}{r_{200,m}} \right)^{a_{1}} \ + \ a_{2} .
\end{equation}

The physical meaning of our parameters is straightforward: a$_0$ represents the normalization of $\alpha_{\rm Turb}$ at r$_{200,m}$, a$_1$ gives the slope of the profile and a$_2$ gives the value of non-thermal support in the cluster center. We notice that \citet{Shi:2014} develop an analytic model to describe the trend of $\alpha_{\rm Turb}$ with the radius. They use three fundamental time scales to develop their model: turbulence dissipation time-scale, t$_{\mathrm{d}}$; the time elapsed between the initial time and the time of observation, (t$_{\mathrm{obs}}$−t$_{\mathrm{i}})$, which characterizes the age of the cluster; and a time-scale characterizing the mass growth rate of the cluster defined by t$_{\mathrm{growth}}$. They define also a turbulence injection efficiency $\eta$, which they constrain to be $\eta \approx 0.5-1$ based on simulations. However, the turbulence injection efficiency is strongly correlated with the slope of the fitting formula, and compared to \citet{Shi:2014} we report a lower injection efficiency, which may also be connected to the role of numerical dissipation of our hydro scheme on small scales.
We also notice that in real systems, and especially at low mass, the turbulence in the core may be dominated by the interplay of cooling and feedback \citep[e.g.,][]{Brighenti02,Bruggen03,Gaspari18}, hence our  a$_2$ may be underestimated. However, we notice that, although our simulations do not include feedback mechanism or cooling, our estimate for a$_2$ is close to the only  available direct spectral measurement from the {\sl Hitomi} \citep[][]{hitomi}. 

\begin{figure}
\includegraphics[width=0.469\textwidth,height=0.5\textwidth]{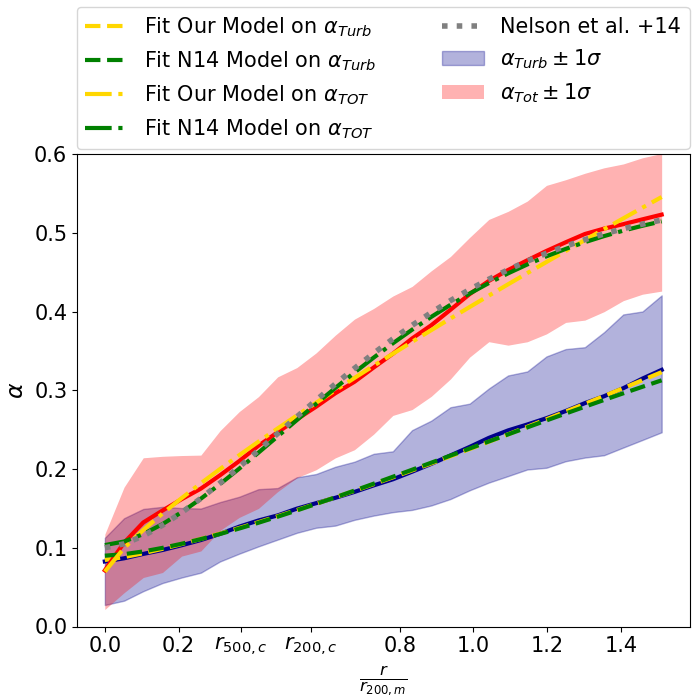}
\caption{Radial profile of median value of $\alpha_{\rm Turb}$ (blue solid line) and $\alpha_{\rm Tot}$ (red solid line). The shadow regions represent the 1$\sigma$ distribution of the sample. The yellow and green dashed lines are the best fits of our model and N14's one on $\alpha_{\rm Turb}$ profile, while the dash-dotted ones are the fits of the models on the $\alpha_{\rm Tot}$ profile. The dotted gray line is the profile obtained in N14.}
\label{fit_our_nelson}
\end{figure}

\citet{2014ApJ...792...25N} present the following analytical fit to the radial distribution of the non-thermal pressure in a dataset of 65 simulated galaxy clusters with masses in a range similar to ours: 
\begin{equation} \label{nelsonmodel}
    \frac{P_{rand}}{P_{tot}}(r) = 1 - A \biggl\{ 1 + \exp \biggl[ - \biggl( \frac{\frac{r}{r_{200,m}}}{B} \biggr)^{\gamma} \biggr] \biggl\} ,
\end{equation}
with best-fit values $A=0.452\pm 0.001$, $B=0.841 \pm 0.008$ and $\gamma=1.628\pm 0.019$. This fit formula is based on three-dimensional gas velocity fields without explicitly filtering out  bulk motions (for details see Sect.~\ref{sec:nelson}). 
The same function also fits our data after filtering, albeit with a slightly higher $\chi^2$ value (see Tab.~\ref{table_chi}). In Fig.~\ref{fit_our_nelson}, we show the different fits overplotted to the median radial profiles of our objects.

\begin{table*}
\begin{tabular}{ccccc}
\multicolumn{5}{c}{Our model} \\
&a$_{0}$&a$_{1}$&a$_{2}$&$\chi^{2}$ \\ \hline \hline
$\alpha_{\rm Turb}$&(1.427$\pm0.002)\cdot10^{-1}$&$1.246\pm0.005$&(8.44$\pm0.02)\cdot10^{-2}$&$10^{-4}$ \\ \hline 
$\alpha_{\rm Tot}$&(3.388$\pm0.009)\cdot10^{-1}$&$0.822\pm0.004$&(6.97$\pm0.08)\cdot10^{-2}$&$0.001$ \\ \hline 
&&&& \\
\multicolumn{5}{c}{N14 model} \\
&A&B&$\gamma$&$\chi^{2}$ \\ \hline \hline
$\alpha_{\rm Turb}$&(4.550$\pm0.003)\cdot10^{-1}$&1.966$\pm0.004$&1.508$\pm0.005$&$2\cdot10^{-4}$  \\ \hline
$\alpha_{\rm Tot}$&(4.483$\pm0.003)\cdot10^{-1}$&0.865$\pm0.002$&1.633$\pm0.006$&$0.046$ \\ \hline
N14&0.45&0.84&1.63 \\ \hline
\end{tabular}
\caption{Parameters and values of $\chi^2$ statistical test for the different formula used to fit the radial behavior of $\alpha_{\rm Turb}$ and $\alpha_{\rm Tot}$. We show also the values of the parameters presented in N14. The errors on the parameters are the values at 3$\sigma$ confidence.}
\label{table_chi}
\end{table*}

From the comparison of the $\chi^2$, it appears that our model yields a better fit to the data than, or as good as,  the model in N14.  The fit suggested by N14 can also fit our data, albeit with different parameters. However, the advantage of our best-fit form is that the fit parameters have a simple physical meaning. 

As we discussed already in \citet{2018MNRAS.481L.120V}, the differences between our results and N14 stem from the different choices in filtering velocities, and the two methods yield formally the same result if no filtering is applied to the 3-dimensional velocity field in our simulations. However, our work suggests that our filtering yields the isotropic part of the turbulent pressure, while filtering out the spurious contribution to the non-thermal pressure support by inward radial motions.

We apply our model also to the profile of $\alpha_{\rm Tot}$ and compare the results with the fit obtained by N14 model. The results are shown in Fig.~\ref{fit_our_nelson} and Tab.~\ref{table_chi}. We notice that our model canl reproduce also the $\alpha_{\rm Tot}$ trend. From the comparison of $\chi^{2}$ we can notice that, as for $\alpha_{\rm Turb}$ case, our model is slightly preferred over the N14 model.

We investigated the possible correlations between the non-thermal pressure and mass, redshift and sparsity of each cluster in our sample. The Sparsity is defined as the ratio between the total mass within r$_{100,c}$ and r$_{200,c}$:
\begin{equation}
   s = \frac{M_{100,c}}{M_{200,c}} .
\end{equation}
For all of these quantities, we divided our samples into three sub-samples which contain the same number of objects. The results are discussed in Appendix~\ref{app:binning}. Here we just note  that our model reproduces the radial behaviors of  $\alpha$ for all subselections of our sample.  The same best-fit parameters are valid for all subsamples, suggesting that no strong dependencies between the non-thermal to total pressure ratio and the cluster mass, the redshift or the mass sparsity 
can be detected in our sample. 

Recently, \citet{eck18published} present some constraints on the non-thermal pressure support in a sample of 12 massive, nearby and mostly relaxed clusters observed in X-rays with {\it XMM-Newton} and in SZ with {\it Planck}. 
They compare the values of the hydrostatic mass, recovered up to r$_{200,c}$ by using a combination of the SZ pressure profile and X-ray based thermodynamical properties \cite[see for details][]{Ettori19}, with mass estimates based on the assumption that hydrodynamical simulations provide the correct baryon fraction distribution in clusters. In the latter estimate, the gas mass is inferred directly from X-ray measurements and the contribution of the mass in stars is evaluated statistically from published work. From the mismatch between the two estimates of the total mass, it is then possible to infer the hydrostatic bias, which turns out to be, on average, consistent with the results obtained by other methods \citep[see][]{Ettori19}. 

If we attribute the origin of this hydrostatic bias to the contribution from a non-thermal pressure component, $P_{\rm nt}(r)$, we can write \cite[e.g.][]{eck18published}:
\begin{equation}
\frac{d}{dr}({\rm P_{th}(r) + P_{nt}(r)}) = - \rho \frac{G M_{T}(<r)}{r^2} ,
\end{equation}
where P$_{\rm th}$ is the thermal pressure component, and $M_{T}$ is the total mass. 
By defining $\alpha(r) = {\rm P_{nt}}(r)/{\rm P_{tot}}(r) = {\rm P_{nt}}(r)/[{\rm P_{nt}}(r)+{\rm P_{th}}(r)]$, the equation above can be rewritten as:
\begin{equation}
M_{T}(<r) = M_{H}(<r) + \alpha(r) M_{T}(<r) - \frac{{\rm P_{th}} r^2}{(1-\alpha) \rho G} \frac{d\alpha}{dr},
\end{equation}
where $M_{H}$ is the hydrostatic mass:
\begin{equation} \label{eq:hydromass}
 M_{H}(i) =  -\frac{\left(\frac{d{\rm P_{th}}}{dr}\right)_i r_i^2}{G \rho_{i-1}}.
\end{equation}
From the equations above and using our radial profiles of total mass  and hydrostatic mass, we can then define $\alpha_{\rm HS}$ at each radius $r$ as:
\begin{equation} \label{alphaeckertformu}
\alpha_{\rm HS} = 1 - \frac{M_{H} + \sqrt[]{(M_{H})^2 - 4 M_{T}  {\rm P_{th}} \frac{r^2}{\rho G} \frac{d\alpha}{dr}}}{2 M_{T} },
\end{equation}
and link it to the parameter $b$, which is usually used in literature to identify the hydrostatic mass bias \citep[e.g.,][]{Salvati2019,2019SSRv..215...25P} 
and defined as
\begin{equation}
    M_{H} = (1-b) M_{T},
\end{equation}
to obtain 
\begin{equation} \label{eq:Bterm}
    b = \frac{\alpha + A}{1 + A},
\end{equation}
where $A$ encloses the pressure's contributions
\begin{equation} 
A = ({\rm P_{th}+P_{nt}}) \frac{d\alpha/dr}{d{\rm P_{th}}/dr}.
\end{equation}

\begin{figure}
$\begin{array}{cc}
\includegraphics[width=0.479\textwidth,height=0.469\textwidth]{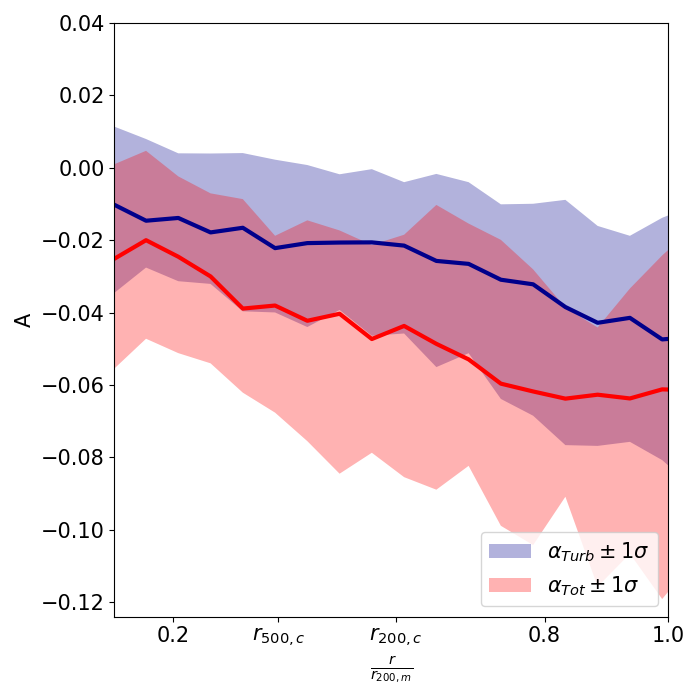}
\end{array}$
\caption{Radial profiles of {\it A} pressure correction term of Eq.~\ref{eq:Bterm}. In blue solid line the profile computed with the radial derivative of $\alpha_{\rm Turb}$,  while in red solid line the profile computed with the $\alpha_{\rm Tot}$. The shadow regions represent the 1$\sigma$ distribution of the data.}
\label{fig:dalfadr}
\end{figure}

We notice that if $\alpha$ is radially constant, then $b=\alpha$. However, $\alpha$ is not generally constant with radius in our sample (see Fig.~\ref{fit_our_nelson}). Hence the $d\alpha/dr$ term plays a small but non-negligible role here. 
As described below, we use both $\alpha_{\rm Turb}$ and $\alpha_{\rm Tot}$ profiles to estimate $d\alpha/dr$ and propagate it to the measurement of $\alpha_{\rm HS}$.
We show in Fig.~\ref{fig:dalfadr} the radial profile of the term {\it A}, where
the different effect of the radial derivative of $\alpha$ allows us to distinguish between the "Turb" and "Tot" cases.

\begin{figure*}
$\begin{array}{cc}
\includegraphics[width=0.995\textwidth]{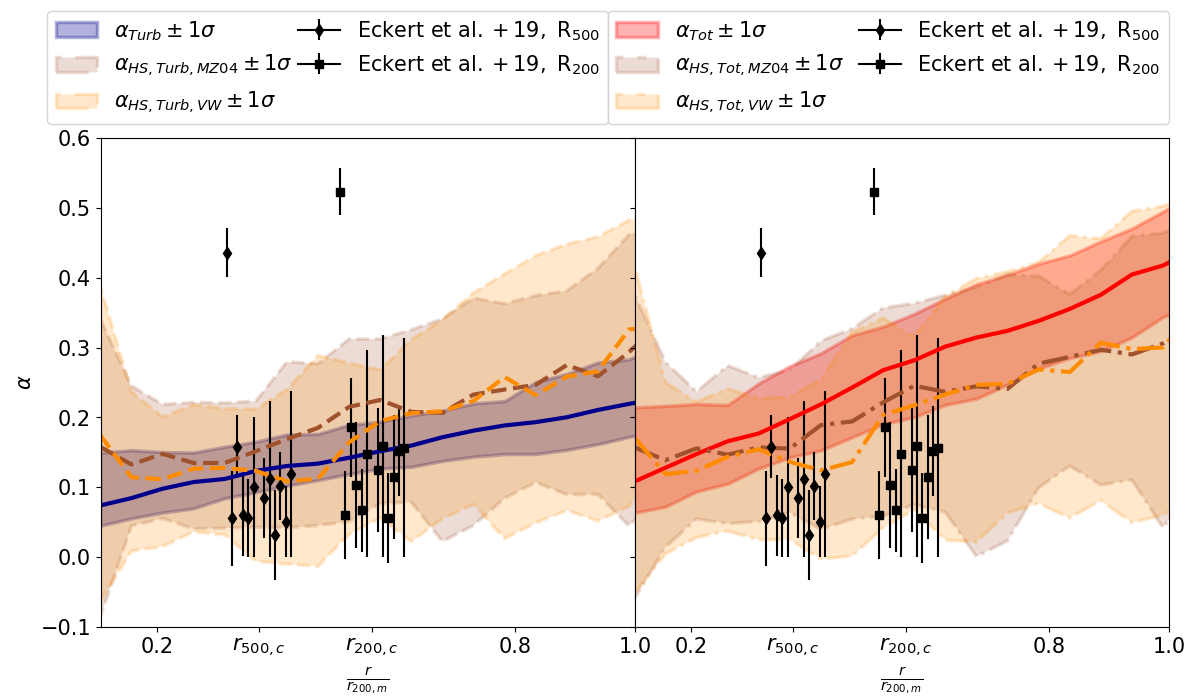}
\end{array}$
\caption{On the left panel, we show the comparison between $\alpha_{\rm Turb}$ (blue solid line) and $\alpha_{\rm HS, Turb}$ computed with volume-weighted profile (orange dashed line) and with the spectroscopic-like temperature (brown dashed line). On the right panel (with the same color-coded legend of the left panel) the result for $\alpha_{\rm Tot}$ against $\alpha_{\rm HS, Tot}$. The shadowed regions give the 1$\sigma$ distribution of the sample for all the profiles. The black points represent the results of \citet{eck18published}.}
\label{volume-mazzotta}
\end{figure*}

We compute the hydrostatic mass $M_{H}$ through the radial derivative of thermal pressure associated with a spectroscopic-like temperature profile, and computed as follows{\footnote{We tested this procedure in idealized (semi-analytically generated) control atmospheres and we verified that our approach is robust to recover the correct mass bias (within a few $\sim 0.01\%$ ) for the typical pressure profile of ICM. Furthermore, we also tested that smoothing our data on scales larger than our spatial resolution ($19.6$ kpc/cell), in order to mimic was can be realistically done by X-ray observations, {\it worsens} the match between the hydrostatic mass and the total mass, especially in 3-dimensional simulated clusters as the smoothing the accuracy with which the internal substructures orbiting in the cluster potential well can be modelled.}}:
\begin{equation}
\left( \frac{d\rm{P_{th}}}{dr}\right)_i = \frac{{\rm P}_{{\rm th}, i-2}-6 \ {\rm P}_{{\rm th}, i-1}+3 \ {\rm P}_{{\rm th}, i}+2 \ {\rm P}_{{\rm th}, i+1}}{6 r_i},
\end{equation}
where P$_{\rm th}$ is the thermal pressure defined as in Eq.~\ref{eq:thermal} and $i$ represents each radial shell.
To limit the contribution from dense, self-gravitating clumps, we use the same masking procedure of Sect.~\ref{sec:kolmogorov}, in order to consider only the thermal pressure exerted by the gas within the cluster.  We also applied a box-averaged smoothing function along ten radial shells at density, thermal pressure and $\alpha_{\rm Turb}$ or $\alpha_{\rm Tot}$ profiles, in order to reduce numerical fluctuations in the profiles. This procedure allows us to obtain smoothing profiles which are more similar to the typical profile obtained in observational works and which are also less affected by spurious numerical effects.

For each radius, we computed  the value of $\alpha_{\rm HS}$ applying the Eq.~\ref{alphaeckertformu}.  The radial derivative of $\alpha$ allows us to define two different $\alpha_{\rm HS}$, already presented above, $\alpha_{\rm HS, Turb}$ and $\alpha_{\rm HS, Tot}$.
To check the dependencies of $\alpha_{\rm HS}$ on physical quantities such as mass, redshift and mass sparsity of clusters, we used the sub-samples analysis presented in Appendix.~\ref{app:binning}. We notice that the variance of the $\alpha_{\rm HS}$ data is larger with respect to the $\alpha$ ones. As for $\alpha$ profiles, also for $\alpha_{\rm HS}$ ones we could not identify any strong correlation with the physical properties of host clusters.

Several factors may influence the comparison between observational results with numerical ones. To test different choices to compute radial  profiles of thermodynamical quantities, in a way similar to what is commonly feasible with observations, we built both a volume-weighted version of $\alpha_{\rm HS}$, as well as one using the "spectroscopic-like" temperature suggested by \citet{mazzotta04} (hereafter MZ04), which approximately takes into account the real spectroscopic response of X-ray detectors in determining the average gas temperature in the ICM. The results are shown in Fig.~\ref{volume-mazzotta}.

We notice that there are no strong differences between the volume-weighted pressure profile and the pressure profile based on the spectroscopic-like temperature. This is consistent with the finding that the hydrostatic mass bias is relatively small in AMR grid-based simulations, while the temperature bias could be large in SPH simulations, owing to the enhanced formation of substructures there \citep{Rasia14}. We also notice how our definition of turbulent motions is more in agreement with the radial trend of $\alpha_{\rm HS}$, mostly in the inner regions that are more accessible to the present X-rays observations. Instead, the $\alpha_{\rm Tot}$ is larger than $\alpha_{\rm HS}$ at any radius. In the following sections, we will use the $\alpha_{\rm HS}$ parameter to easily compare our results to the ones in \citet{eck18published}, as show in the plots (black points with errorbars).

Similarly to \citet{2018MNRAS.481L.120V}, but extended in the present study to the full set of simulated clusters, the scatter in the simulations is typically larger than the one observed in real cluster data, with the exception of A2319 \citep[][]{2018A&A...614A...7G}, which probably comes from the intrinsic difference in the two samples: the X-COP sample contains by selection mostly relaxed clusters, while our sample contains a larger variety of objects. We will further comment on this issue in Sect.~\ref{sec:conclusions}. 
Despite this promising average agreement between these two samples, in the next section we investigate the caveats which may lead to a mismatch between observational estimates of $\alpha_{\rm HS}$ and the underlying presence of turbulent motions in single objects.

It shall be noticed that all approaches are in better agreement in the center of clusters, although the scatter is large for the distribution of $\alpha_{\rm HS}$. Interestingly, the central median values of $\alpha_{\rm Turb}$ and $\alpha_{\rm Tot}$ for the sample are in the $5-8\%$ range, which is reasonably close to  the most recent estimates from the centre of the Perseus cluster, i.e. $\sim 2-6\%$ \citep[][]{hitomi18}. This trend is well explained by \citet{Lau17} and \citet{Bourne17}.

\subsection{On the relation between turbulence and hydrostatic bias} \label{sec:turbuandhydro}

\begin{figure*}
\includegraphics[width=0.469\textwidth,height=0.5\textwidth]{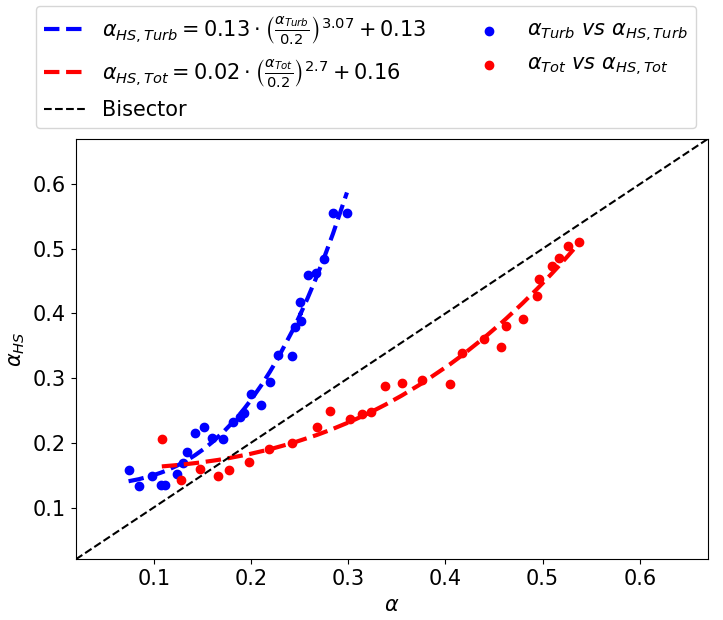} 
\includegraphics[width=0.469\textwidth,height=0.5\textwidth]{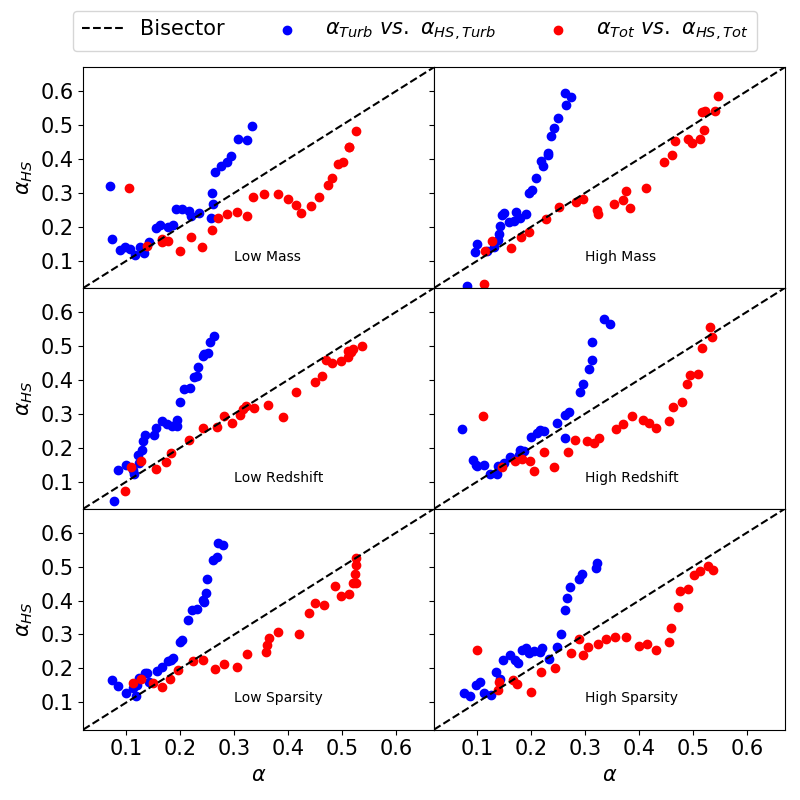}
\caption{Relation between $\alpha$ and $\alpha_{\rm HS}$ computed as described in Sect.~\ref{sec:turbuandhydro}. In the left side the whole sample, where the blue points are obtained from $\alpha_{\rm Turb}$ and $\alpha_{\rm HS, Turb}$, while the red ones are obtained from $\alpha_{\rm Tot}$ and $\alpha_{\rm HS, Tot}$. The dashed lines are the fits made as described in Sect.~\ref{sec:turbuandhydro}, where the coefficients are given in the legend. The black dashed line is the bisector. In the right panel there are six sub-samples, as described in Sect.~\ref{sec:turbuandhydro}. Top panels: low mass on the left, high mass on the right; central panels: low redshift on the left, high redshift on the right; bottom panels: low sparsity on the left, high sparsity on the right. Also in these plots the blue point represent the "Turb" quantities, while the red dots represent the "Tot" ones. The dashed lines are the plots' bisectors.}
\label{fig:scatter_radial}
\end{figure*}

To investigate the relation between turbulence and hydrostatic bias in a more systematic way, 
we compare the radial dependence of $\alpha$ and $\alpha_{\rm HS}$. 
We consider the median values of entire sample of $\alpha$ and $\alpha_{\rm HS}$ in each radial shell. The results are shown in left panel of Fig.~\ref{fig:scatter_radial}. 

\begin{figure*}
\includegraphics[width=0.469\textwidth,height=0.469\textwidth]{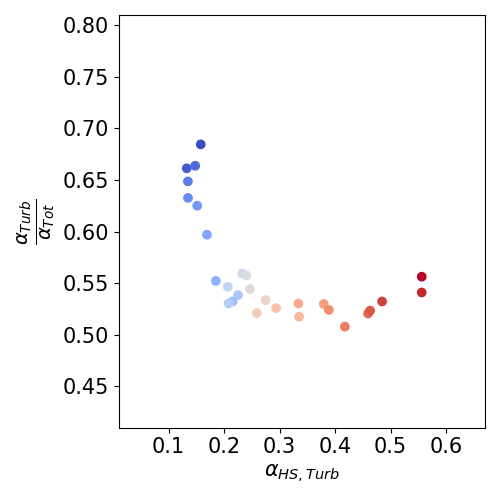} 
\includegraphics[width=0.469\textwidth,height=0.469\textwidth]{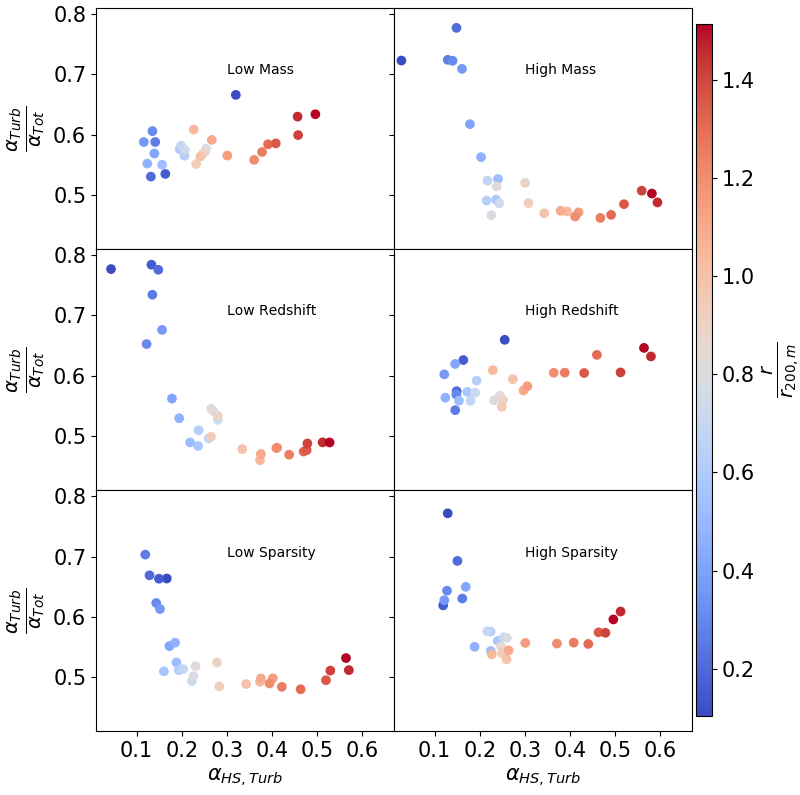}
\caption{Relation between the ratio $\alpha_{\rm Turb}/\alpha_{\rm Tot}$ and $\alpha_{\rm Hs, Turb}$. In the left side the whole sample, while in the right one there are six sub-samples, as described in Sect.~\ref{sec:turbuandhydro}. Top panels: low mass on the left, high mass on the right; central panels: low redshift on the left, high redshift on the right; bottom panels: low sparsity on the left, high sparsity on the right. The colors represent the distance from the cluster's center.}
\label{fig:scatter_turb}
\end{figure*}

If turbulence and hydrostatic bias were perfectly correlated, 
the data would closely follow the bisector of the plot. 
Instead, we find that their relation is well described by a function $\alpha_{\rm HS} = a_0 \times (\alpha/0.2)^{a{_1}} + a_2$ (see Fig.~\ref{fig:scatter_radial}). This suggests that in general both our definition of turbulence and the N14 model do not trace in a 1-to-1 relation the hydrostatic bias. 
From Fig.~\ref{fig:scatter_radial}, it also appears that the N14 definition of turbulence is much closer to the 1-to-1 relation between turbulence and hydrostatic bias. Considering that the N14 model includes not only turbulent motions but also the kinetic pressure associated to bulk motions, we conclude that the latter one represents an important ingredient for the mass bias,
especially when large values of $\alpha_{\rm HS}$ are considered 
(which happens most often at larger radii).

We divide further our sample in two mass, redshift or sparsity sub-samples (see  right panel of Fig.~\ref{fig:scatter_radial}). We observe that, as opposed to the analysis of the complete sample, $\alpha_{\rm Tot}$ is able to reproduce the 1-to-1 relation with $\alpha_{\rm HS, Tot}$ in some sub-samples. This confirm that $\alpha_{\rm Tot}$ is a more accurate tracer of hydrostatic bias compared to our definition of $\alpha_{\rm Turb}$. This is supported by the plot in Fig.~\ref{fig:scatter_turb} where we shown the ratio between $\alpha_{\rm Turb}$ and $\alpha_{\rm Tot}$ against $\alpha_{\rm HS}$ (here we used $\alpha_{\rm HS, Turb}$ but the results are almost the same for the $\alpha_{\rm HS, Tot}$ case). 
Combining the information from Fig.~\ref{fig:scatter_radial} and Fig.~\ref{fig:scatter_turb}, we notice that our definition of turbulent motions give a value of $\alpha$ which is always a fraction of the value obtained from the N14 filtering technique, $\sim 50-70\%$ from case to case. From Fig.~\ref{fig:scatter_turb} we also notice that less is the hydrostatic bias, greater is the support from purely turbulent motions to non-thermal pressure. Indeed, the higher is $\alpha_{\rm HS}$, the lower is the ratio $\alpha_{\rm Turb}$ against $\alpha_{\rm Tot}$. This ratio has also a dependence with distance from the cluster's core. Indeed, greater is the radius and lower is the ratio. Therefore, the role of the turbulent motions in the non-thermal pressure support is less important in cluster outskirts than in the innermost regions.
\begin{figure*}
$\begin{array}{cc}
\includegraphics[width=0.469\textwidth,height=0.469\textwidth]{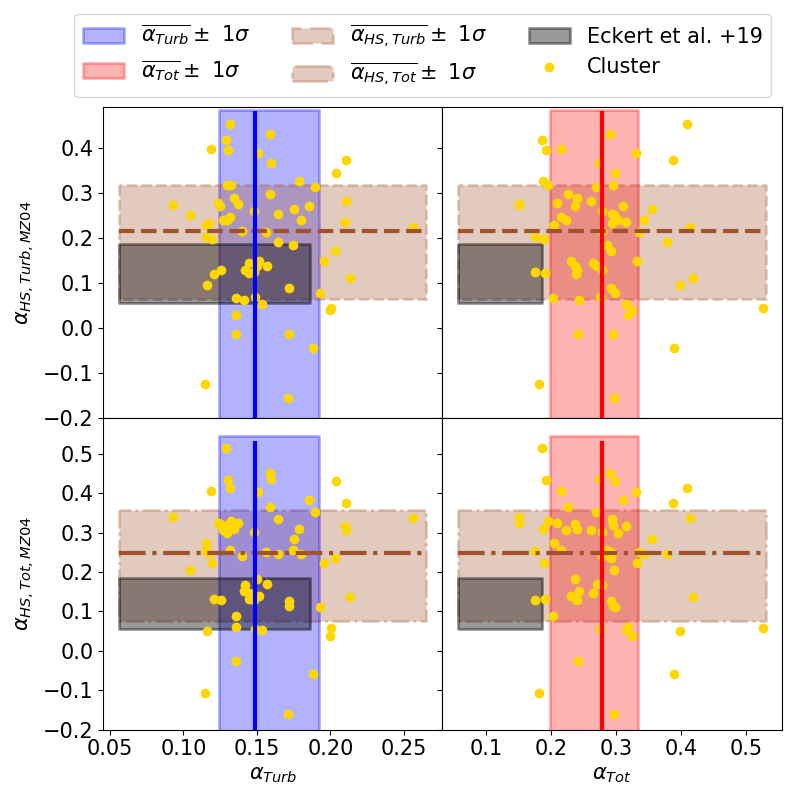} 
\includegraphics[width=0.469\textwidth,height=0.469\textwidth]{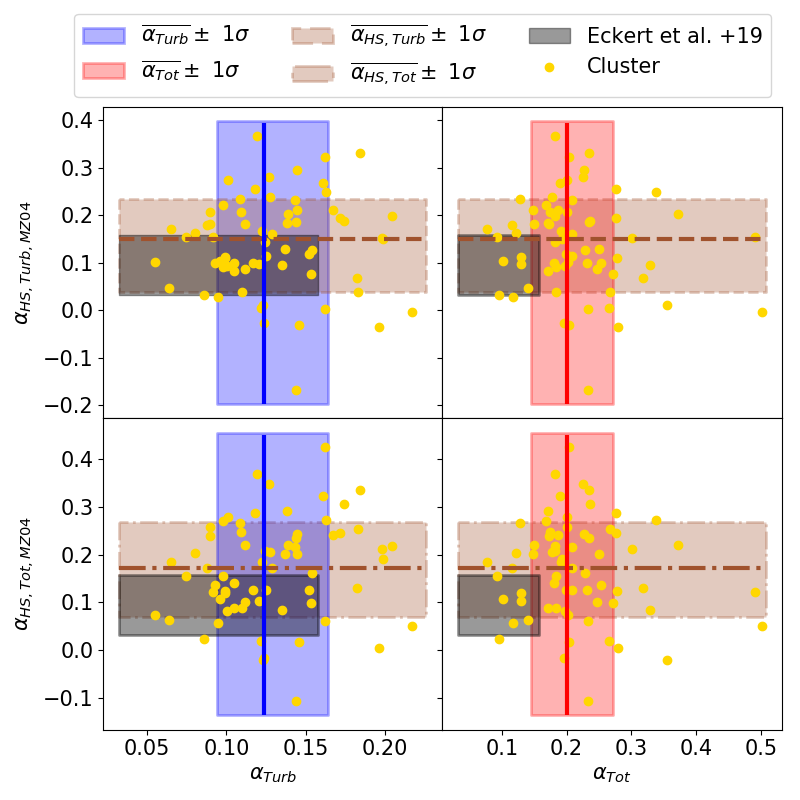} 
\end{array}$
\caption{Comparison between median $\alpha_{\rm Turb}$ (blue solid line) or $\alpha_{\rm Tot}$ (red solid line)  and $\alpha_{\rm HS, Turb}$ (brown dashed line) or $\alpha_{\rm HS, Tot}$ (brown dash-dotted lines) at r$_{200,c}$ (left panels) and r$_{500,c}$ (right panels). The shadow regions represented 1$\sigma$ variance. The black shadowed regions represented the range values presented by \citet{eck18published}, while the yellow dots refer to  singles clusters (we notice that one cluster of the sample is found to be well outside the plot limits and is discarded given its very peculiar shape).}
\label{alfaturbvsalfahs}\end{figure*}
We made also a sub-samples analysis of these relations and the results are shown in the right panel of  Fig.~\ref{fig:scatter_turb}. We notice that for all the sub-samples the above conclusions still hold. In case of low mass or high redshift clusters,  we also notice that the ratio $\alpha_{\rm Turb}$ against $\alpha_{\rm Tot}$ is almost constant with the hydrostatic bias. 

We conclude that the relation between turbulence and hydrostatic bias can be statistically detected only with a typical number of $\geq 20$ objects, at least in the mass range probed by our sample.
On the other hand, when applied to single objects the analysis presented above appears not suitable to give accurate correction factors for the hydrostatic bias, as we will show in the next chapter.

\section{Discussion} \label{sec:discussion}

We can finally study the relation between the non-thermal pressure, the turbulence we identify in our data and the X-ray derived proxy \citep[e.g.][]{eck18published}.
To address this, we compute the values of $\alpha_{\rm HS}$ and $\alpha$ at the radii r$_{200,c}$ and r$_{500,c}$ in each object. The results are shown in Fig.~\ref{alfaturbvsalfahs}

At first glance, there are almost no correlations between the two proxies ($\alpha$ and $\alpha_{\rm HS}$) for the non-thermal pressure support, even if the two distributions span a similar range of values, and also are in the same ballpark of the XCOP results discussed in \citet{eck18published}. We also notice that while $\alpha$ is defined as a positive quantity by construction, $\alpha_{\rm HS}$ can be measured with negative values (in $\sim$10\% of the sample) due to local fluctuations in the reconstructed thermodynamical profiles that are responsible of the scatter in the distribution of the estimated $\alpha_{\rm HS}$ at a given radius.
\begin{figure*}
\includegraphics[width=0.469\textwidth,height=0.469\textwidth]{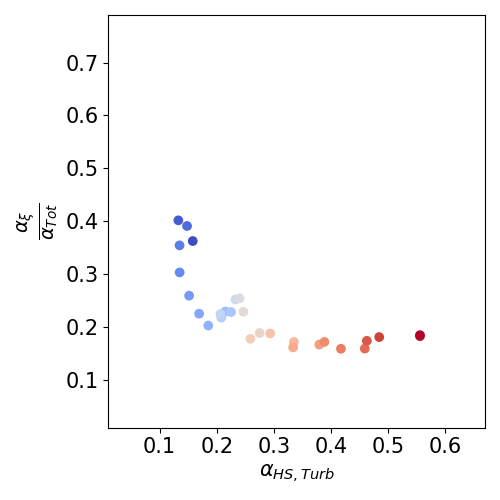} 
\includegraphics[width=0.469\textwidth,height=0.469\textwidth]{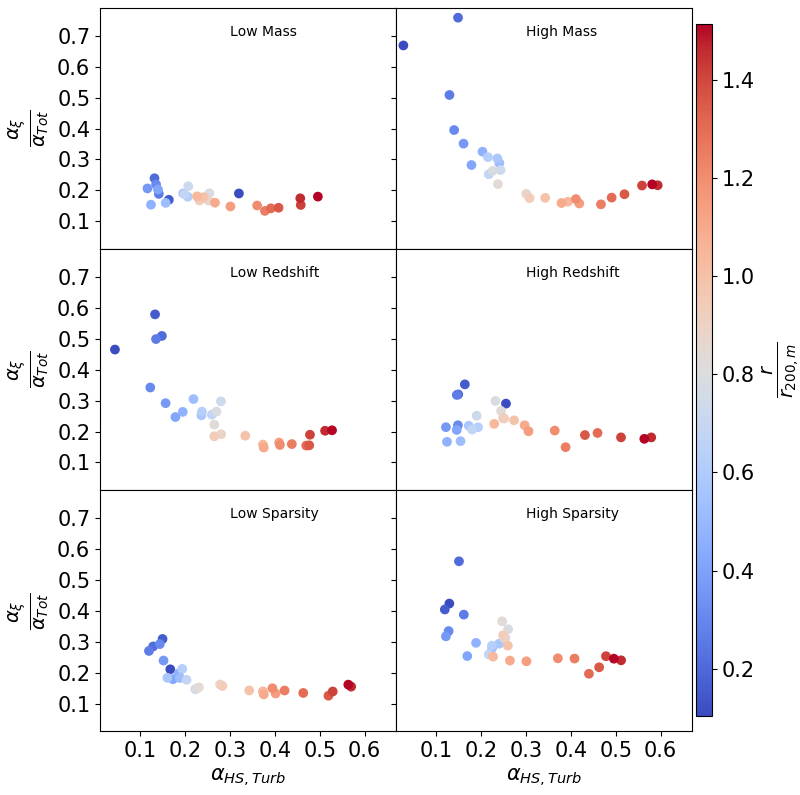}
\caption{Relation between the ratio $\alpha_{\xi}$ against $\alpha_{\rm Tot}$ and $\alpha_{\rm Hs, Turb}$. In the left side the whole sample, while in the right one there are six sub-samples, as described in Sect.~\ref{sec:turbuandhydro}. Top panels: low mass on the left, high mass on the right; central panels: low redshift on the left, high redshift on the right; bottom panels: low sparsity on the left, high sparsity on the right. The colors represent the distance from the cluster's center.}
\label{fig:scatter_acceler}
\end{figure*}
This introduces crucial problems for the $\alpha_{\rm HS}$ estimate. First,  the spherical symmetry and the coincidence between the centre of the gas pressure and of the gravitational mass (and between gas and dark matter densities) are often violated for systems which had only a little time to relax. Gas substructures are also more prominent, as they are often found in their first crossing of the ICM. This also leads to an ICM with a multiphase structure, also correlated with the crossing of shocks. In summary, most of the assumptions on which the hydrostatic equilibrium analysis is based are violated at high-z, while on the other hand the above factors little affect our estimate of $\alpha$, because through our filtering procedure the measure of turbulence is local and do not rely in assumptions of symmetry or isothermality. Interestingly, the above problems should also play an important role for the mass modelling of high-z galaxy clusters in real observations \citep[e.g.][]{2006MNRAS.365..509M,2011ApJ...737...59J,2018MNRAS.474.2635S}. 

To further investigate the physical origin of the deviations from hydrostatic equilibrium, we added a few important dynamical proxies to characterize the dynamics of our systems.More specifically, we computed  the radial profile of gas acceleration, and derived the residual acceleration from gas motions which are out of equilibrium in the presence of mergers, following  \citet{2016ApJ...827..112B}. 
While \cite{2016ApJ...827..112B} could directly access the acceleration values of single smoothed-particle-hydrodynamics (SPH) particles from the hydrodynamical solver, we rely on the post-processing of  Eulerian data, taking the derivative of two close timesteps. 
We defined the gravitational acceleration in each radial shell as:
\begin{equation}
g(r) = -\frac{G M_{T}}{r^2} ,
\end{equation}
while the residual gas acceleration is computed by first taking the radial velocity in each cell, and then reconstructing the radial profile of this quantity for every selected snapshot. To define the residual gas acceleration in the radial direction, we take the difference in each radial shell, between two snapshots, $\delta(r)$ as:  
\begin{equation}
\delta(r)=\frac{Vr(t_2)-Vr(t_1)}{(t_2 - t_1)}. 
\end{equation}
In order to follow  a convention consistent with \citet{2016ApJ...827..112B}, we defined an acceleration term consistent with the one extracted from their SPH simulations:
\begin{equation}
H(r)=g(r)+\delta(r). 
\end{equation}
From the above we can thus introduce a factor, $\delta_{HE}$, which compensates for the residual gas radial acceleration by motions which are not in equilibrium with the gravitational pull of the cluster:  
\begin{equation} \label{eq:deltahe}
\delta_{HE}(r) = \frac{g(r)}{H(r)} - 1. 
\end{equation}
From this definition we notice that when the gas is in hydrostatic equilibrium $\delta_{HE}$ is equal to 0. 
Finally, as in \citet{2016ApJ...827..112B}  we define $\xi_r$:
\begin{equation}
\xi_r = | \ \alpha_{\rm HS}(r) - \delta_{HE}(r) \ |,
\end{equation}
which allows us to consider at the same time the contributions given at the hydrostatic bias from the acceleration terms and the term obtained from $\alpha_{\rm HS}$.  In the Appendix~\ref{app:acceleration}, we show in more detail how this procedure can well highlight the presence of out-of-equilibrium conditions in the ICM; in particular we can relate large residual acceleration terms to the local gas conditions in the proximity of powerful shock waves crossing the cluster (see Fig.C1-C2). 
 We can thus conclude that these terms are likely to be crucial for an accurate estimate of hydrostatic mass, but also that they are hard to model for a perfect correction of the hydrostatic mass bias, in  most realistic cases.

We attempted to incorporate residual gas accelerations in the previous modelling of the hydrostatic bias, by defining $\alpha_{\xi}$ as the fraction of the total pressure due to purely radial accelerations:
\begin{equation}
\alpha_{\xi}(r) = | \ \overline{\delta_{HE}}(<r) \ |
\label{eq:alpha-xi}
\end{equation}
where $\overline{\delta_{HE}}$ is defined starting from the Eq.~\ref{eq:deltahe} and it represents the median value of the contributions of residual radial accelerations within the radius {\it r}. In Fig.~\ref{fig:scatter_acceler} we show the behavior of the ratio $\alpha_{\xi}/\alpha_{\rm Tot}$ against $\alpha_{\rm HS}$.  

From the left panel of Fig.~\ref{fig:scatter_acceler} we notice that the contribution of radial accelerations to the non-thermal pressure support span from $\sim$40\% in the innermost regions of the cluster to $\sim$15\% in the outskirts. Two effects are likely at play here. First, strong acceleration terms originate from expanding shocks, which are typically launched during mergers starting from the core cluster region (or close to it). Therefore, when a cluster is disturbed by ongoing merger activity, such terms can be more significant in the core regions than in more external ones, due to their larger volume filling fraction there.
Additionally, we expect an irreducible level of spurious radial acceleration at the scale of cluster cores, due the fact that in perturbed systems (i.e. with many substructures in the core etc) it is not always trivial to exactly track the position of the centre as a function of time, which may introduce noise in our procedure to measure $\delta_{HE}(<r)$ above.  

As for the turbulent contributions, which is discussed in Sect.~\ref{sec:turbuandhydro}, also for the radial accelerations one, the higher is the hydrostatic bias and lower is the contribution. The sub-sample analysis revealed that for high masses and low redshift samples, in the innermost regions of the clusters where the hydrostatic bias is lower, the contribution of radial accelerations seems to be higher than for low masses and high redshifts.  

\begin{figure*}
\includegraphics[width=0.469\textwidth,height=0.469\textwidth]{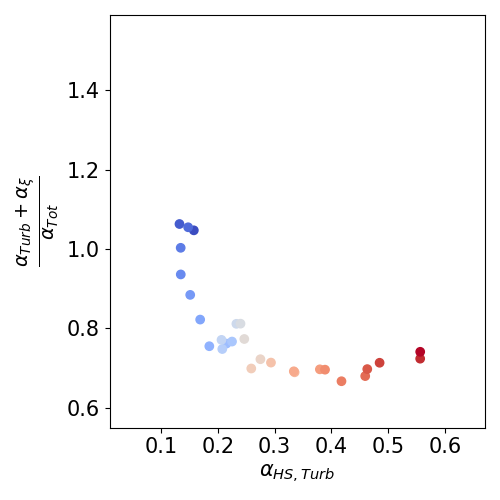} 
\includegraphics[width=0.469\textwidth,height=0.469\textwidth]{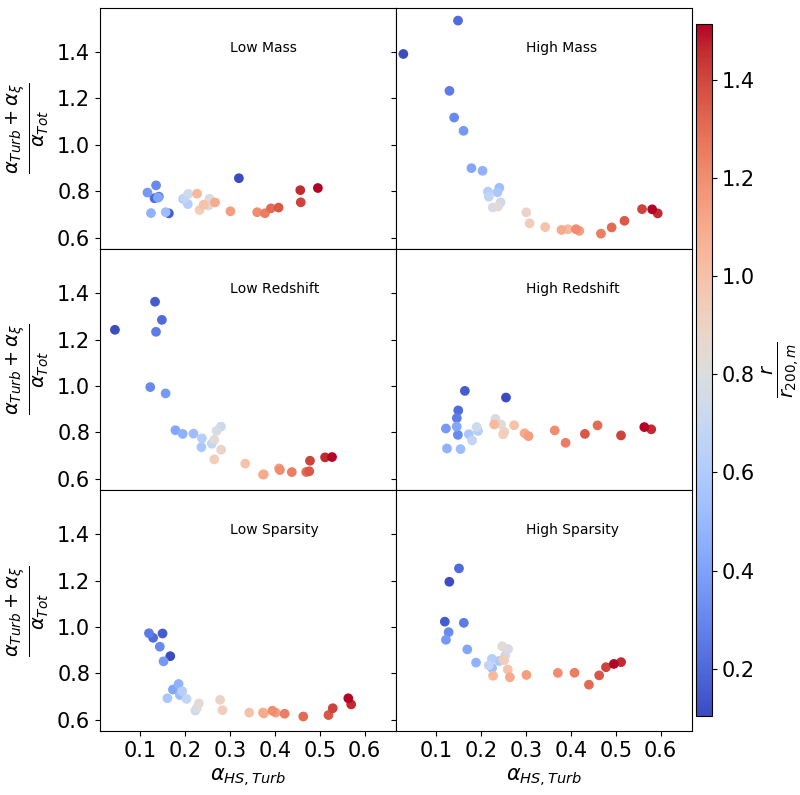}
\caption{Relation between the ratio $\alpha_{\rm Turb}+\alpha_{\xi}$ against $\alpha_{\rm Tot}$ and $\alpha_{\rm Hs, Turb}$. In the left side the whole sample, while in the right one there are six sub-samples, as described in Sect.~\ref{sec:turbuandhydro}. Top panels: low mass on the left, high mass on the right; central panels: low redshift on the left, high redshift on the right; bottom panels: low sparsity on the left, high sparsity on the right. The colors represent the distance from the cluster's center.}
\label{fig:scatter_turbacceler}
\end{figure*}

Combining the contribution from turbulence and radial accelerations, we obtained the results shown in Fig.~\ref{fig:scatter_turbacceler}. In the clusters outskirts, where the hydrostatic bias is higher, turbulence and radial accelerations are not able to completely trace the hydrostatic bias. Due to the filtering techniques, the residual 30\% of the hydrostatic bias is generated by the bulk motions, which are filtered out from our filtering technique, but they are considered in the filter proposed by N14. However, in the innermost regions the hydrostatic bias is completely described by the combination of turbulence and radial accelerations contributes.

We performed the same sub-sample analysis used above. For low masses and high redshift clusters, the combination of turbulence and radial accelerations is not still enough to account for the hydrostatic bias. On the other hand, for high masses, low redshift or low sparsity objects, where the hydrostatic bias is lower, turbulence and radial accelerations trace the hydrostatic bias. However, even then, turbulence and radial accelerations are not still sufficient to explain the hydrostatic bias.

\section{Conclusions} \label{sec:conclusions}

The hydrostatic bias plays an important role in limiting the use of galaxy clusters as cosmological probes \citep[e.g.][and references therein for recent reviews]{2019SSRv..215...25P,Salvati2019}. 
Using a recent sample of 68 galaxy clusters simulated at high resolution with the cosmological code {\enzo} \citep[][]{va17turbo,wi17b,2018MNRAS.481L.120V}, we  have shown that cosmological simulations combined with a sophisticated filtering of turbulent motions can in principle predict the same value of the hydrostatic mass bias inferred from the combination of X-ray and SZ observations \citep[e.g.][]{eck18published}, relying on the simplistic hydrodynamical view of the inviscid intracluster medium. 

This work improves on our previous works on the subject in several respects.
Firstly, we designed a procedure to extract a larger sample of objects and conduct larger statistical studies, by extracting multiple snapshots of the same objects with a sufficiently large time separation to consider them as dynamically independent clusters (see Sect.~\ref{sec:cosmoselect}). This resulted in a total sample of 68 clusters. 
Secondly, we improved the iterative small-scale filtering techniques used in  \citet{2018MNRAS.481L.120V}, by linking the tolerance parameter in our multi-scale iterative analysis of turbulence to the expected increase of velocity with scale, following Kolmogorov theory (Sect.~\ref{sec:kolmogorov}). 

We thus extracted the three-dimensional distribution of turbulent velocities from which we computed the non-thermal contribution of pressure, $\alpha_{\rm Turb}$, as a function of cluster-centric distance. We also compared this definition of turbulence using different filtering technique, specifically to the one proposed in \citet{2014ApJ...792...25N}, we defined as $\alpha_{\rm Tot}$. 

To compare to observations, we also computed the hydrostatic mass bias, defined via the parameter $\alpha_{\rm HS}$, and we studied the possible factors leading to a mismatch with the corresponding $\alpha_{\rm Turb}$ or $\alpha_{\rm Tot}$ measured for the same systems.

Our main results can be summarized as follows:
\begin{itemize} 
\item We derived fitting formula for the radial profiles of $\alpha_{\rm Turb}$ and of $\alpha_{\rm Tot}$  (Eq.~\ref{eq:fitalfaturb}).
Such fitting formulae reproduce the data of our simulations, both for the complete sample and for the different sub-samples which we studied. We found that the three free parameters of our fiting model can be easily related to the physics of the ICM (see Sect.~\ref{sec:fitting} for a detailed explanation and the necessary bibliographic references). 
\item The average non-thermal pressure support in our sample is in agreement with the recent X-ray observational campaign by \citet{eck18published}, both at  r$_{500,c}$ and r$_{200,c}$, albeit with a large scatter which is probably due to a larger variety of simulated dynamical states,  compared to the X-COP sample used by \citet{eck18published} (Fig.~\ref{volume-mazzotta}).  Different choices in filtering  generate very different non-thermal pressure support \citep[see][]{2018MNRAS.481L.120V}, and we show that the other filtering techinque for turbulence  yields values of non-thermal pressure support which are $\sim 2-3$  times larger. Our analysis of subsamples in mass, redshift and sparsity sub-samples do not evidence particular trends. Despite having a different selection bias for clusters, the comparison between the non-thermal pressure support obtained by \citet{eck18published} yields rather similar results.
\item The two approaches do not differ much in the center of clusters, and when applied to our cluster sample the median central value of the non-thermal pressure support is well in line with the most recent estimates from the centre of the Perseus cluster, i.e. $\sim 2-6\%$ \citep[][]{hitomi18}. This result is in agreement with recent results by \citet{Lau17} and \citet{Bourne17}, who also highlighted the plausible impact of  "cosmic weather" driven by cluster dynamics, such as mergers and/or gas sloshing, in contributing to the budget of turbulent motions on the small scales of cluster cores. 
\item Following on \citet{eck18published}, we computed $\alpha_{\rm HS}$ for our objects, defined as in Eq.~\ref{alphaeckertformu}. When averaged over the sample, the median $\alpha_{\rm HS}$ is not far from the values obtained by X-ray observations, i.e. $\alpha_{\rm HS} \sim 10-20\%$ at $r_{\rm 500,c}$ and $r_{\rm 200,c}$, albeit with a large scatter. The hydrostatic mass bias does not show strong dependences on the cluster mass or the cluster dynamical status (see Sect.~\ref{sec:fitting} and Sect.~\ref{sec:discussion} for details).
\item We compared the values of $\alpha_{\rm Turb}$, $\alpha_{\rm Tot}$  and $\alpha_{\rm HS}$ at different radii both for the median of our sample (see Fig.~\ref{fig:scatter_radial}) and for single objects (Fig.~\ref{alfaturbvsalfahs}).
In general, we found no strong correlation between these parameters in the latter case, whereas we find a strong correlation when the median values of the sample are considered.
\item We studied the ratio between $\alpha_{\rm Turb}$ and $\alpha_{\rm Tot}$ against $\alpha_{\rm HS}$, finding that 
(i) there is a minimal contribution from $\alpha_{\rm Turb}$ to $\alpha_{\rm Tot}$ of about 50\% at every radius;
(ii) when $\alpha_{\rm HS} \geq 0.2$, at large radii, half of the hydrostatic mass bias is due to kinetic energy components which are not purely turbulent, with an increasing contribution by bulk or large-scale flows. 
These results do not depend on the mass, redshift and sparsity of the halo (see Sect.~\ref{sec:turbuandhydro}).
\item Finally, we assess the contribution of radial accelerations studying the the ratio between $\alpha_{\xi}$ and $\alpha_{\rm Tot}$ against $\alpha_{\rm HS}$ (see Sect.~\ref{sec:discussion} and the bibliographic references therein).  
This contribution spans from 40\% to 15\%, from cluster's core to the outskirts. We conclude that the sum of the contributions given by turbulence and radial accelerations could completely explain the observed hydrostatic bias in the innermost regions of the clusters, but far from the center, the combination of turbulence and radial accelerations' terms reaches $\sim$70\% (see Fig.~\ref{fig:scatter_turbacceler}).
\end{itemize}

In summary, we find that a well-defined correlation between turbulence and hydrostatic bias emerges from the average properties of a sample of clusters. 
On the other hand, if we focus on single objects, the internal dynamics of each cluster makes it impossible to find a simple relation between the level of turbulence and its hydrostatic bias. 

This irreducible bias stems from episodic radial acceleration terms, related to merger activity (Appendix.~\ref{app:acceleration}), confirming earlier results \citep[][]{2014ApJ...792...25N,2016ApJ...827..112B}. 
Shocks and, more generally, non-thermal phenomena observed in the radio band are expected to trace important out-of-equilibrium conditions in the intracluster medium \citep[][]{2014IJMPD..2330007B,2019SSRv..215...16V}. 

In this respect, future observations, e.g., with LOFAR and SKA may be able to provide important clues to the presence of significant non-thermal pressure in dynamically disturbed systems. Thanks to numerical simulations, such out-of-equilibrium conditions may be linked to the presence of turbulent motions, as observations have also begun to establish the quantitative link between observed radio power in radio halos to the turbulent energy budget of the ICM, inferred from the amount of fluctuations in X-ray surface brightness \citep[e.g.][]{eck17}.
This will be key for the cosmological use of galaxy clusters in future X-ray surveys (e.g. with eRosita, see for example \citealt{2018MNRAS.480..987Z}). Furthermore, deep exposures of clusters in X-rays will enable the calibration of the $\alpha_{\rm HS}$-$\alpha$ relation by measuring spectroscopically the level of gas turbulence (e.g. with ATHENA, see for example \citealt{2018A&A...618A..39R} and \citealt{Vazza2019}). 

\section*{acknowledgements}

We thank our anonymous reviewer for the very valuable scientific feedback on the first version of our paper.
We thank E. Rasia and V. Biffi for valuable scientific discussions, that helped us with the physical interpretation of our results. 
The cosmological simulations described in this work were performed using the {\enzo} code (http://enzo-project.org), which is the product of a collaborative effort of scientists at many universities and national laboratories. We gratefully acknowledge the {\enzo} development group for providing extremely helpful and well-maintained on-line documentation and tutorials. 

M.A. and F.V. acknowledge financial support from the European Union's Horizon 2020 program under the ERC Starting Grant "MAGCOW", no. 714196. 

T.J. acknowledges financial support from the US NSF through grant AST1714205.

S.E. acknowledges financial contribution from the contracts ASI 2015-046-R.0, ASI-INAF n.2017-14-H.0, and from INAF "Call per interventi aggiuntivi a sostegno della ricerca di main stream di INAF".

The PhD grant supporting M.A. is co-funded from INAF and ERC Starting Grant "MAGCOW", no. 714196.

\bibliographystyle{mnras}
\bibliography{franco.bib}

\appendix

\section{Testing the relation between filtering scale and non-thermal pressure}
\label{app:f_test}
\begin{figure}
\includegraphics[width=0.469\textwidth]{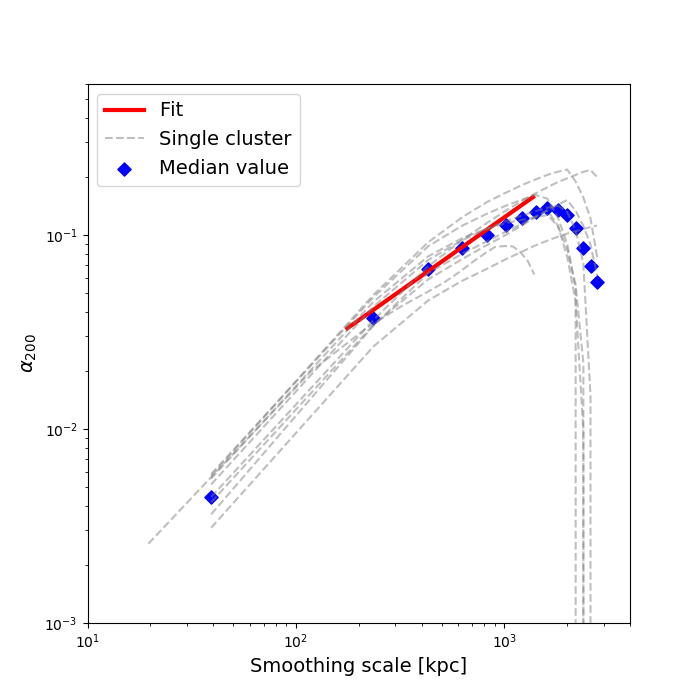}
\caption{Relation between smoothing scales and $\alpha_{200}$ for the sample of clusters at $z=0$. The blue points represent the median values of $\alpha_{200}$ at different smoothing scales for the used sample of clusters. The solid red line gives the power-law fit obtain explained in Sect.~\ref{sec:kolmogorov}.}
\label{kolmo_med}
\end{figure}
As detailed in the main text (Sec.\ref{sec:kolmogorov}) our iterative scheme to measure the local turbulent velocity field requires to predict the increase of the local rms velocity as a function of the smoothing scale, $w$, i.e .$\delta v^2 \propto w^f$. 
In Kolmogorov theory such exponent should be $f=2/3$, 
yet we tested to what extent the standard Kolmogorov theory of turbulence applies to our data, and studied the relation between the value of $\alpha$ measured at the reference radius of r$_{200,c}$ ($\alpha_{200}$) as a function of different fixed smoothing scale.
If we apply the standard relation between rms velocity and turbulent scale (e.g. $\sigma^2$ $\propto$ L$^{\frac{2}{3}}$) in the stationary subsonic turbulent regime described by Kolmogorov theory, we expect that the relation should approximately follow:
\begin{equation} 
\label{kolmoeq}
\alpha_{200} = {a} \cdot {w}^{{f}} ,
\end{equation}
where $w$ is the value of smoothing scale in physical quantities and $a$ and $f$ are the parameters obtained from Kolmogorov's theory \citep[][]{Kolmogorov1941}. The expected value for $f$ is close to $\frac{2}{3}$ for stationary and subsonic turbulence. However, the ICM is not such an idealized environment because of density stratification, self-gravity and non-stationary flow patterns, which can lead to deviations from 2/3. 
Fig.~\ref{kolmo_med} shows the pressure ratio, $\alpha_{200}$, versus the smoothing scale computed of our set of clusters at $z=0$, after computing the local turbulent velocity field for increasing smoothing scales, $w$.
The trend is measured to be  very similar across our sample, and can be fitted by a unique power-law.

We fit the data to Eq.~\ref{kolmoeq} and obtain $a \simeq 6 \cdot 10^{-3}$ and $f \simeq 0.77$, which is the fiducial value we adopted to stop the iterations in our method in the main paper, as in Eq.4. 
The value for $f$ is reasonably close to 2/3 and is consistent with the fact that the power spectra of the velocity field in simulated galaxy clusters are typically steeper than Kolmogorov's slope because of the stratified cluster atmosphere \citep[][]{va11turbo}.
Only the scales below $\sim$100 kpc show hints of a steepening, which may partially be ascribed to numerical dissipation in the PPM scheme, which is expected to dampen the velocities on scales close to a few times the spatial resolution (e.g, \citealt{pw94}). 
For scales larger than $\sim 8$ times the numerical resolution ($\geq$200 kpc) these effects do not occur and the relation between $\alpha$ and the smoothing scale is well fitted by Kolmogorov's spectrum. Since a number of physical and numerical effects may affect the dynamics of the turbulent flow on $<$100-200 kpc, with these simulations it is hard to tell the different effects apart. In the following, we focus mostly on the dynamics of turbulence on scales $>$100 kpc, which are also the ones that dominate the non-thermal pressure support. 
On scales greater than $\sim$1 Mpc, the spectra show a drop where the peak of Kolmogorov spectrum is reached. 
The exponent $f$ in Eq.~\ref{kolmoeq} is calculated in the inertial range of Kolmogorov spectrum, from $\sim$200 to $\sim$ 800 kpc, so we can use this value for the multi-scale adaptive filtering.

\section{Sub-samples analysis of $\alpha$ radial profiles} \label{app:binning}

\begin{figure*}
\includegraphics[width=0.995\textwidth,height=1.25\textwidth]{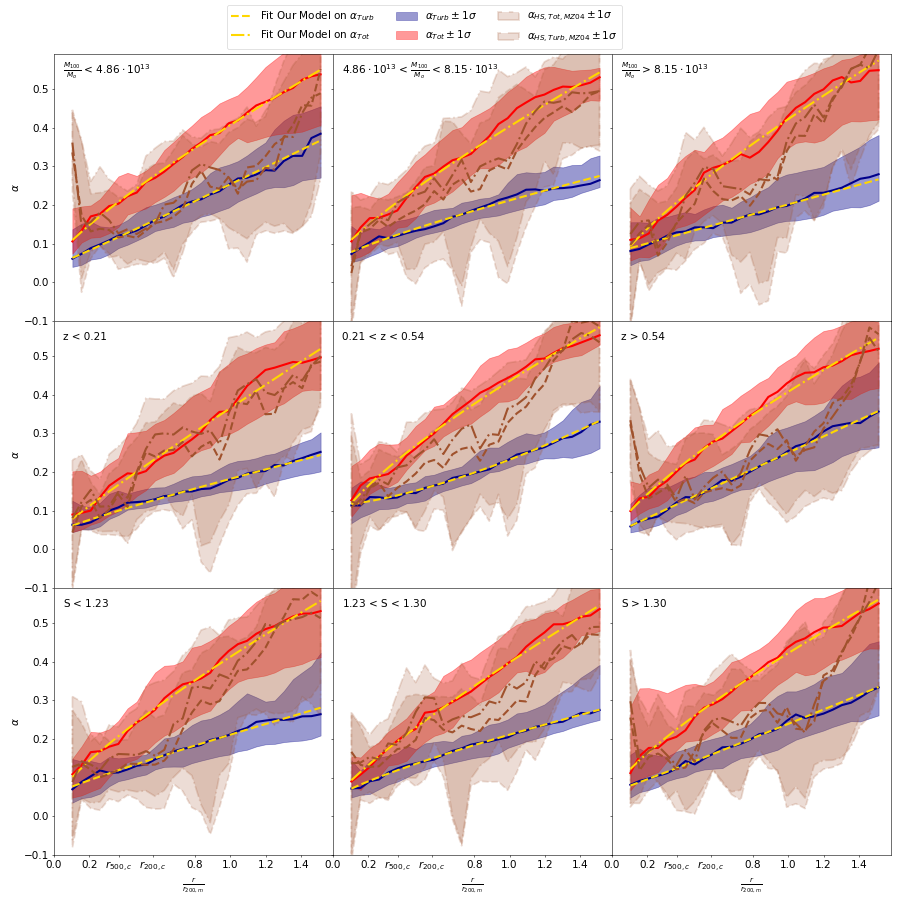}
\caption{Radial profile of median value of $\alpha_{\rm Turb}$ (blue solid line) and $\alpha_{\rm Tot}$ (red solid line) for different bins of mass (top panels), redshift (central panels) and sparsity (bottom panels). The yellow dashed lines are the fits of our model on $\alpha_{\rm Turb}$ profile, while the dash-dotted ones are the fits of the model on the $\alpha_{\rm Tot}$ profile. The brown dashed and dash-dotted lines are the profile of median $\alpha_{\rm HS, Turb}$ and $\alpha_{\rm HS, Tot}$ computed as described in Sect.~\ref{sec:fitting}. The shadow regions represent the 1$\sigma$ distribution of the sub-sample. }
\label{fig:fit_all}
\end{figure*}

As described in Sect.~\ref{sec:fitting}, we adopt our model as the best fitting function both for $\alpha_{\rm Turb}$ and $\alpha_{\rm Tot}$.
In Fig.~\ref{fig:fit_all} we shown the radial profiles of $\alpha$ and $\alpha_{\rm HS}$. Here the profiles of $\alpha_{\rm HS}$ are computed using a the spectroscopic-like definition of the temperature profile. As already discussed in Sect.~\ref{sec:fitting}, the differences between a spectroscopic-like profile and a volume-weighted one are as small that they allow us to use either of the two profiles.
The sub-samples used in this appendix are built to have the same number of objects in all the bins. The values which allow us this selection are shown in Tab.~\ref{tab:tablesubsamples}.

Both from Fig.~\ref{fig:fit_all} and Tab.~\ref{tab:tablesubsamples}, we conclude that our fitting formula well reproduce the radial behaviour of $\alpha_{\rm Turb}$ and $\alpha_{\rm Tot}$ at any sub-samples. For $\alpha_{\rm Turb}$ we observe a slightly decrease at any radii with an increase of mass or a decrease of redshift. However, no other strong dependencies are observed between the radial profiles and the quantities used for the selection. We notice also that $\alpha_{\rm Turb}$ shows the lowest scatter at any radii, while both $\alpha_{\rm HS, Turb}$ and $\alpha_{\rm HS, Tot}$ are affected by high scatters.

\begin{table*}
\begin{tabular}{cccccccc}
\multicolumn{2}{c}{Sample}&&a$_{0}$&a$_{1}$&a$_{2}$&$\chi^{2}$ \\ \hline \hline
&\multirow{2}{*}{$M_{100}/M_{\odot}<4.86 \cdot 10^{13}$}&$\alpha_{\rm Turb}$&(2.100$\pm0.005)\cdot10^{-1}$&1.041$\pm$0.003&(0.430$\pm0.003)\cdot10^{-1}$&0.010 \\
    &&$\alpha_{\rm Tot}$&0.351$\pm0.001$&0.825$\pm$0.004&0.054$\pm0.001$&0.008 \\[2ex]
\multirow{2}{*}{Mass}&\multirow{2}{*}{$4.86 \cdot 10^{13}<M_{100}/M_{\odot}<8.15 \cdot 10^{13}$}&$\alpha_{\rm Turb}$&(1.582$\pm0.001)\cdot10^{-1}$&0.815$\pm$0.002&(0.532$\pm0.001)\cdot10^{-1}$&0.034 \\
    &&$\alpha_{\rm Tot}$&(3.467$\pm0.003)\cdot10^{-1}$&0.821$\pm$0.001&(0.589$\pm0.001)\cdot10^{-1}$&0.035 \\[2ex] 
&\multirow{2}{*}{$M_{100}/M_{\odot}>8.15 \cdot 10^{13}$}&$\alpha_{\rm Turb}$&(1.261$\pm0.004)\cdot10^{-1}$&1.003$\pm$0.007&(0.755$\pm0.005)\cdot10^{-1}$&0.014 \\
    &&$\alpha_{\rm Tot}$&$0.395\pm0.001$&0.786$\pm$0.004&0.027$\pm0.001$&0.036 \\ \hline
&\multirow{2}{*}{$z<0.21$}&$\alpha_{\rm Turb}$&(1.398$\pm0.001)\cdot10^{-1}$&0.884$\pm$0.001&(0.424$\pm0.001)\cdot10^{-1}$&0.024 \\
    &&$\alpha_{\rm Tot}$&(3.271$\pm0.005)\cdot10^{-1}$&0.929$\pm$0.004&(0.386$\pm0.003)\cdot10^{-1}$&0.030 \\[2ex]
\multirow{2}{*}{Redshift}&\multirow{2}{*}{$0.21<z<0.54$}&$\alpha_{\rm Turb}$&(1.186$\pm0.001)\cdot10^{-1}$&1.488$\pm$0.002&(1.130$\pm0.001)\cdot10^{-1}$&0.023 \\
    &&$\alpha_{\rm Tot}$&(3.875$\pm0.008)\cdot10^{-1}$&0.743$\pm$0.002&(0.474$\pm0.008)\cdot10^{-1}$&0.040 \\[2ex]
&\multirow{2}{*}{$z>0.54$}&$\alpha_{\rm Turb}$&(2.113$\pm0.003)\cdot10^{-1}$&0.998$\pm$0.004&(0.380$\pm0.004)\cdot10^{-1}$&0.012 \\
    &&$\alpha_{\rm Tot}$&0.373$\pm0.001$&0.769$\pm$0.004&0.034$\pm0.001$&0.027 \\ \hline
&\multirow{2}{*}{$s<1.23$}&$\alpha_{\rm Turb}$&(1.537$\pm0.003)\cdot10^{-1}$&0.903$\pm$0.004&(0.573$\pm0.003)\cdot10^{-1}$&0.023 \\
    &&$\alpha_{\rm Tot}$&(3.730$\pm0.006)\cdot10^{-1}$&0.793$\pm$0.002&(0.392$\pm0.001)\cdot10^{-1}$&0.019 \\[2ex]
\multirow{2}{*}{Sparsity}&\multirow{2}{*}{$1.23<s<1.30$}&$\alpha_{\rm Turb}$&(1.674$\pm0.002)\cdot10^{-1}$&0.791$\pm$0.002&(0.435$\pm0.003)\cdot10^{-1}$&0.012 \\
    &&$\alpha_{\rm Tot}$&(3.587$\pm0.003)\cdot10^{-1}$&0.836$\pm$0.001&(0.390$\pm0.003)\cdot10^{-1}$&0.023 \\[2ex]
&\multirow{2}{*}{$s>1.30$}&$\alpha_{\rm Turb}$&(1.761$\pm0.003)\cdot10^{-1}$&1.046$\pm$0.005&(0.629$\pm0.004)\cdot10^{-1}$&0.014 \\
    &&$\alpha_{\rm Tot}$&(0.361$\pm0.002)\cdot10^{-1}$&0.785$\pm$0.006&0.061$\pm0.002$&0.014 \\ \hline
\end{tabular}
\caption{Parameters and values of $\chi^2$ statistical test for our model applied to mass, redshift and mass sparsity sub-samples of our data. The errors on the parameters are the values at 3$\sigma$ confidence.}
\label{tab:tablesubsamples}
\end{table*}

\section{On the relation between hydrostatic bias and radial acceleration} \label{app:acceleration}
\begin{figure*}
$\begin{array}{cc}
\includegraphics[width=0.469\textwidth,height=0.469\textwidth]{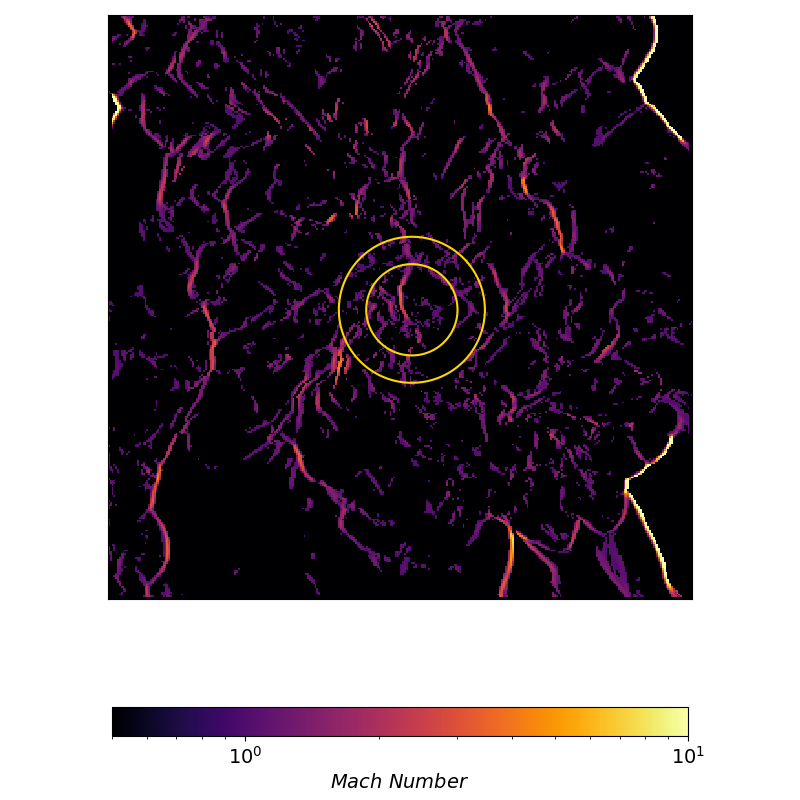} &
\includegraphics[width=0.469\textwidth,height=0.469\textwidth]{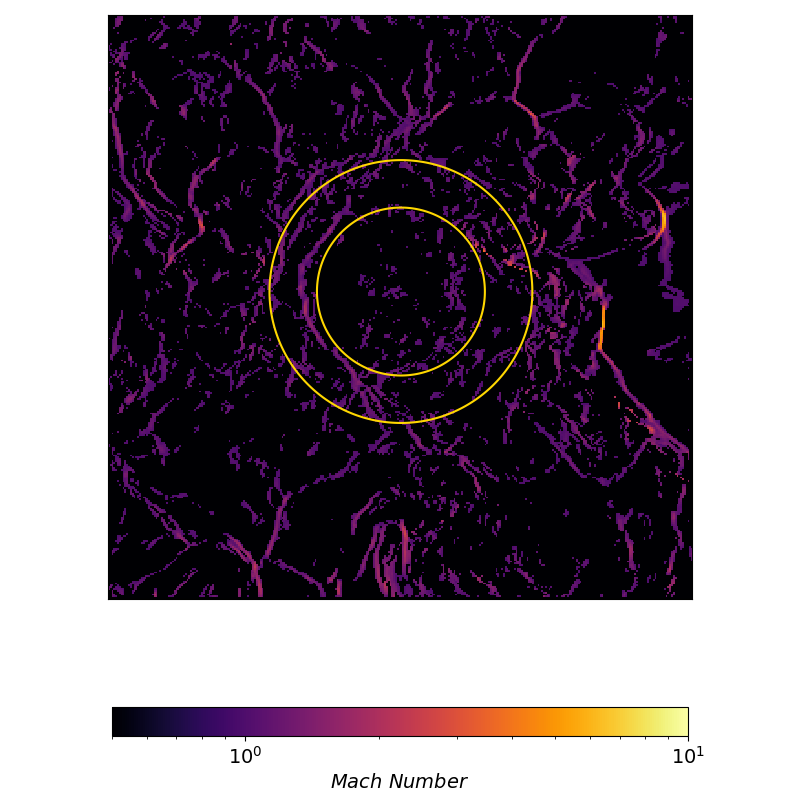}  
\end{array}$
\caption{Maps of shock Mach number in a slice through the  center for cluster IT90\_0 at $z\simeq0.15$ (left panel), and through the center of IT92\_0 at $z\simeq0.07$ (right panel). The inner circle shows the location of r$_{500,c}$, while the outer one shows  r$_{200,c}$.}
\label{massvsshock_map}
\end{figure*}
\begin{figure*}
$\begin{array}{cc}
\includegraphics[width=0.469\textwidth,height=0.65\textwidth]{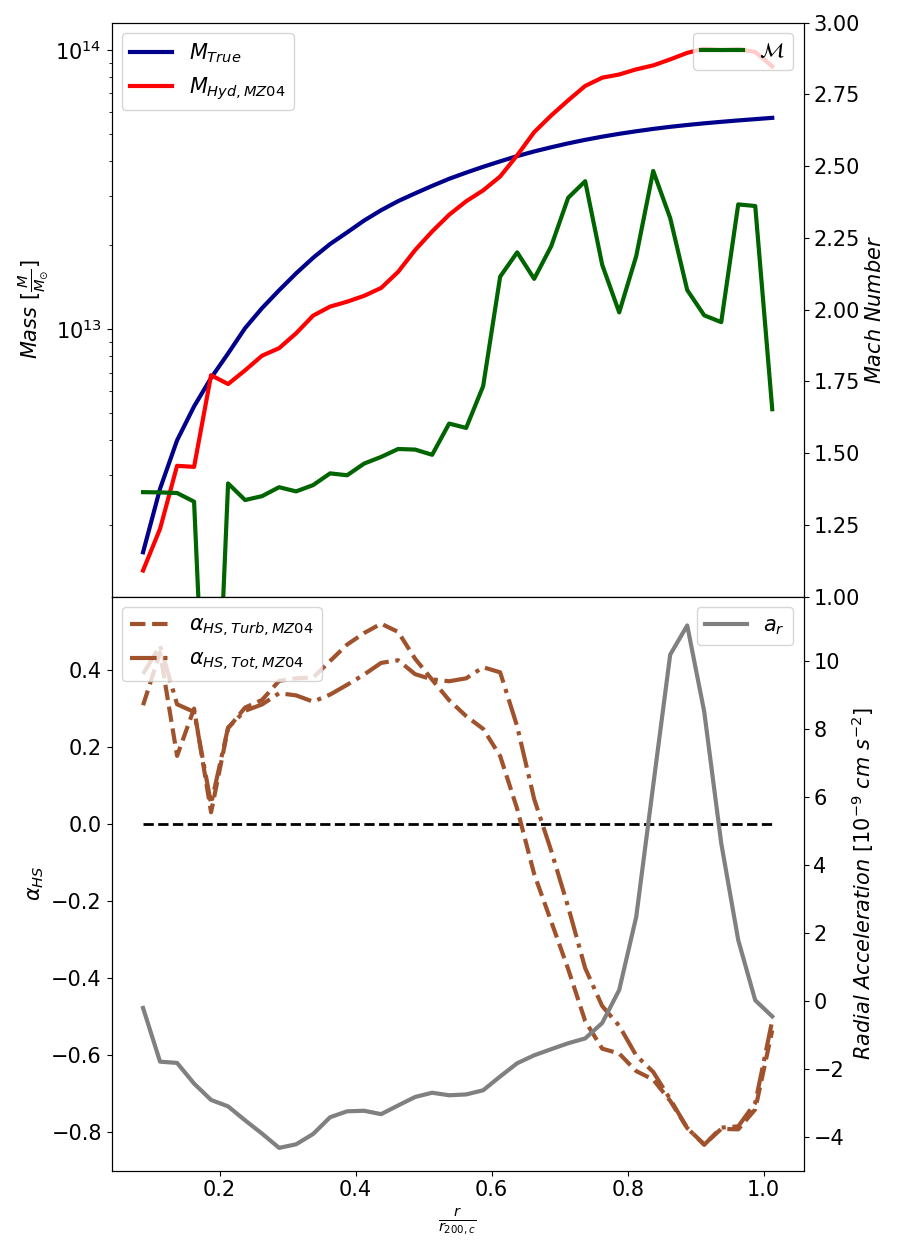} &
\includegraphics[width=0.469\textwidth,height=0.65\textwidth]{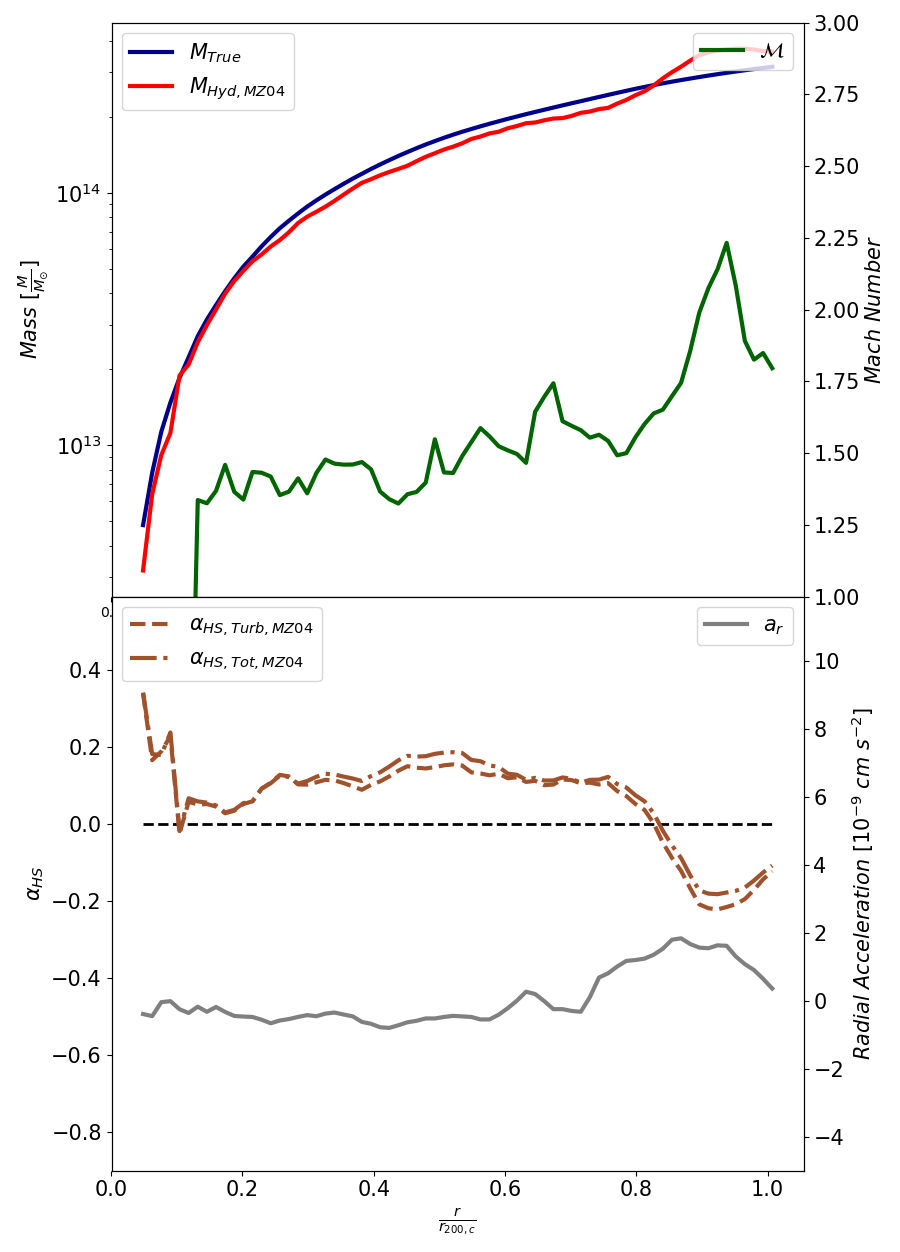}  
\end{array}$
\caption{Total mass profile (blue solid line), hydrostatic mass profile (red solid line) and median Mach Number profile (green solid line) in the top panel, and radial acceleration (blue solid line) and $\alpha_{\rm HS, Turb}$ (brown dashed line) and $\alpha_{\rm HS, Tot}$ profiles (brown dash-dotted line) in the bottom panel, for IT90\_0 at $z\simeq0.15$ (left panel) and IT92\_0 at $z\simeq0.07$ (right panel).}
\label{massvsshock_prof}
\end{figure*}

\begin{figure*}
$\begin{array}{cc}
\includegraphics[width=0.469\textwidth,height=0.469\textwidth]{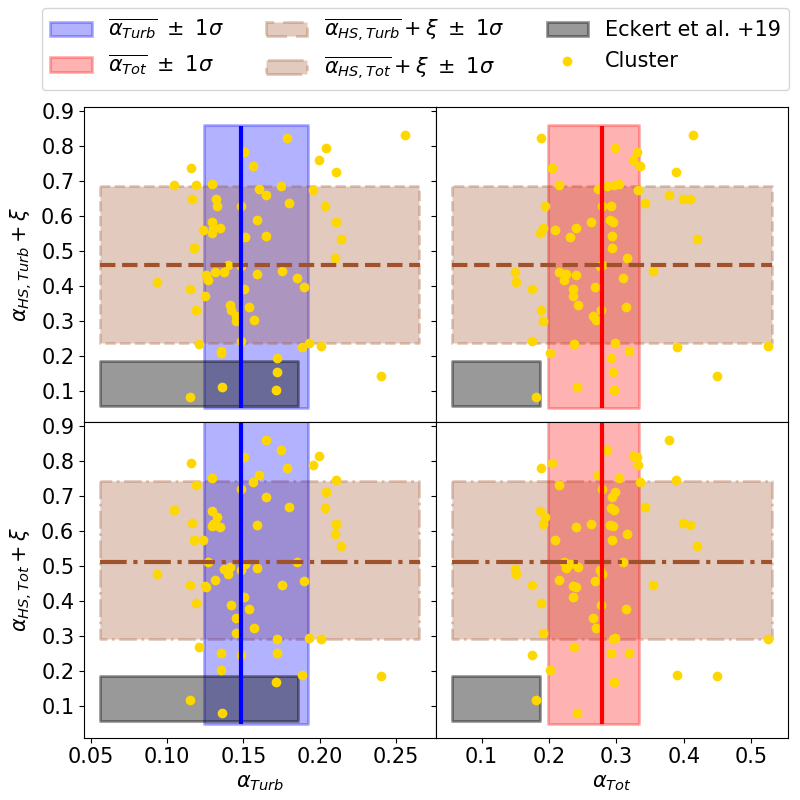} &
\includegraphics[width=0.469\textwidth,height=0.469\textwidth]{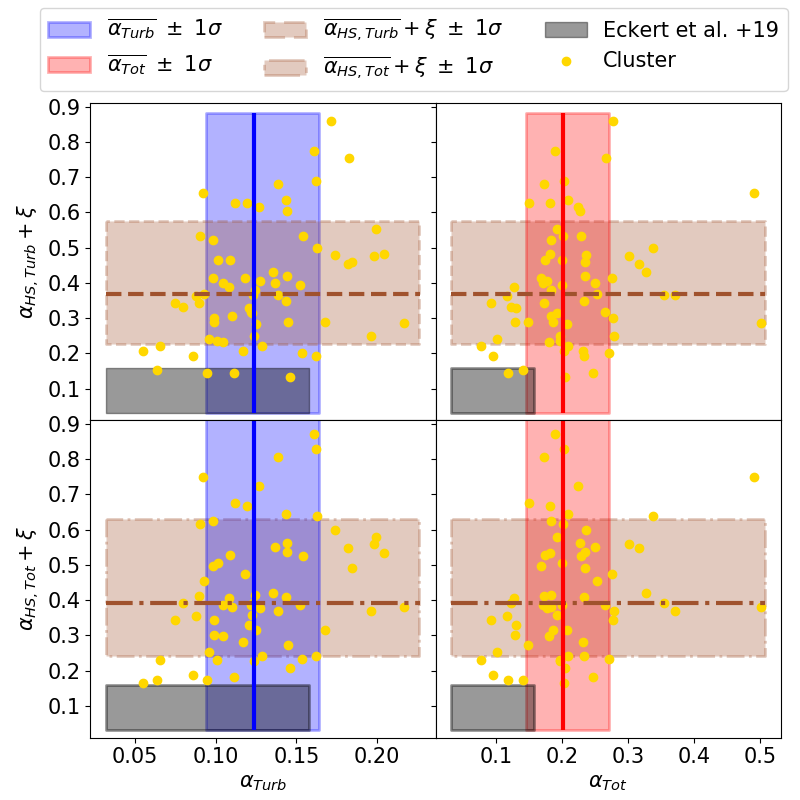}
\end{array}$
\caption{Comparison between median $\alpha_{\rm Turb}$ (blue solid line) or $\alpha_{\rm Tot}$ (red solid line)  and $\alpha_{\rm HS, Turb}$ (brown dashed line) or $\alpha_{\rm HS, Tot}$ (brown dash-dotted lines) corrected for the radial acceleration term $\xi$, at r$_{200,c}$ (left panels) and r$_{500,c}$ (right panels). The shadow regions represented 1$\sigma$ variance. The black shadow regions represented the range values inferred with X-ray observations by \citet{eck18published}, while the yellow dots refer to single clusters.}
\label{alfaturbvsalfahscorr}
\end{figure*}

The presence of shocks in the ICM is important for their dynamical equilibrium, since passage of a shock provides a thrust, usually in the outward direction. This effect generates a radial acceleration of the gas that could affect the computation of the hydrostatic mass, mimicking an excess of thermal pressure if the hydrostatic equilibrium is (wrongly) imposed on the structure \citep{nelson14b}. \\
As written in Sect.~\ref{sec:discussion}, we apply the same formalism presented in \citet{2016ApJ...827..112B} on our clusters.
We quantify the amount of departure from the hydrostatic equilibrium in each shell through the median value of $\xi_r$ within the shell, $\xi$. 
As an example, in Fig.~\ref{massvsshock_map} we show the central slice of Mach Number, which allows us to identify shocks sweeping the clusters volume at a given epoch. In left panel we can see a wide  $\cal{M} \approx \rm 3$ shocks in the inner part of cluster IT90\_0 at the epoch of $z \simeq 0.15$, while in right slight through cluster IT92\_0 at $z \simeq 0.07$ there are no relevant shocks inside  r$_{200,c}$. We can therefore expect in the first case a stronger departure from equilibrium, following the gasdynamical acceleration downstream of the shock wave.  These trends are well captured in Fig.~\ref{massvsshock_prof}, which gives the radial behavior of total mass (blue solid line), hydrostatic mass (red solid line) and $\cal{M_{\rm w}}$ (green solid line). In the bottom panels of Fig.~\ref{massvsshock_prof} we also show the radial profiles of $\alpha_{\rm HS}$ and of the radial acceleration term. 

From there we can quantify how shocks in the inner parts of the cluster influence the hydrostatic mass. There is a strong correlation between the maximum values of the Mach Number and a {\it negative}  hydrostatic mass bias, meaning that the total mass that would be inferred through a standard hydrostatic equilibrium analysis would be larger than the total (true) mass, as shown by the radial trend of $\alpha_{\rm HS}$. 
These behaviors are also observed for the more relaxed cluster in the right panel, but only in the regions close to r$_{200,c}$, where a shock front is visible in the right panel of Fig.~\ref{massvsshock_map}. 
Therefore, shocks introduce an additional term which one must consider when inferring non-thermal pressure from the hydrostatic mass bias. 
We remark that such behaviours in the radial profile would hardly be detected in realistic X-ray analysis of observed clusters, because observations are usually fitted through a (smooth) Navarro-Frank-White profile, which cannot produce such a  sharp increase in the hydrostatic mass profile. \\
Thanks to the analysis of radial acceleration terms, we used $\xi$ as correcting factor of hydrostatic bias. We compared the $\alpha$ terms with the sum of $\alpha_{\rm HS}$ plus $\xi$. The goal of this comparison is to obtain a closer relation between the turbulent proxy $\alpha$ and the hydrostatic bias counterpart $\alpha_{\rm HS}$. The results are shown in Fig.~\ref{alfaturbvsalfahscorr}. \\
From the comparison between Fig.~\ref{alfaturbvsalfahscorr} and Fig.~\ref{alfaturbvsalfahs}, we notice that the negative terms of hydrostatic bias are not present anymore. However, we do not find a strong correlation between turbulence and hydrostatic bias yet. \\

\end{document}